\documentstyle[12pt,twoside]{article}  

\newfont{\twlvmsb}{msbm10 scaled\magstep1}
\newfont{\ninemsb}{msbm9}
\newfont{\sixmsb}{msbm6}
\newfam\msbfam
\textfont\msbfam=\twlvmsb
\scriptfont\msbfam=\ninemsb
\scriptscriptfont\msbfam=\sixmsb
\catcode`\@=11
\def\Bbb{\ifmmode\let\next\Bbb@\else
  \def\next{\errmessage{Use \string\Bbb\space only in math 
mode}}\fi\next}
\def\Bbb@#1{{\Bbb@@{#1}}}
\def\Bbb@@#1{\fam\msbfam#1}
\newfont{\largeeufm}{eufm10 scaled\magstep4}
\newfont{\twlveufm}{eufm10 scaled\magstep1}
\newfont{\elveufm}{eufm10 at 11pt}
\newfont{\teneufm}{eufm10}
\newfont{\nineeufm}{eufm9}
\newfam\eufam
\textfont\eufam=\twlveufm
\scriptfont\eufam=\nineeufm
\def\frak{\ifmmode\let\next\frak@\else
\def\next{\errmessage{Use \string\frak\space only in math mode}}\fi\next}
\def\frak@#1{{\fam\eufam{{#1}}}}
\catcode`@=12

\catcode `\@=11
\@addtoreset{equation}{section}
  
\def\theequation{\arabic{section}.\arabic{equation}} 

\def\section{\@startsection {section}{1}{\z@}{3ex plus 1ex minus
.2ex}{1.2ex plus .2ex}{\normalsize\bf}}

\def\subsection{\@startsection{subsection}{2}{\z@}{3ex plus 1ex minus
.2ex}{1.2ex plus .2ex}{\normalsize\bf}}

\def\thebibliography#1{\section*{{
				    \rm REFERENCES}\@mkboth
 {BIBLIOGRAPHY}{BIBLIOGRAPHY}}\list
 {[\arabic{enumi}]}{\settowidth\labelwidth{[#1]}\leftmargin\labelwidth
 \advance\leftmargin\labelsep\usecounter{enumi}}
 \def\newblock{\hskip .11em plus .33em minus -.07em}
 \sloppy\clubpenalty4000\widowpenalty4000
 \sfcode`\.=1000\relax}

\catcode `\@=12

\title{\vspace*{1.7cm} \normalsize\bf 
CANONICAL STRUCTURE OF CLASSICAL FIELD THEORY 
IN THE POLYMOMENTUM PHASE SPACE\thanks{
{\sl PACS classification:} 03.50, 02.40}
\thanks{
{\sl  AMS classification:} 70 G 50, 58 F 05, 53 C 80, 
81 S 10}
\thanks{
{\sl Keywords}: classical field theory, 
Poincar\'e--Cartan form,
De Donder--Weyl theory, Hamiltonian formalism, polysymplectic form, 
multivector fields, differential forms, 
\mbox{Schouten--}Nijenhuis bracket, Poisson bracket, 
Gerstenhaber algebra.}
}

\author{\vspace{1.7cm} \sc Igor V. Kanatchikov\thanks{e-mail: 
kai@fuw.edu.pl} \vspace{-1.7cm}\\    
\small Laboratory of Analytical Mechanics and Field Theory \vspace{-0.2cm}\\ 
\small Institute of Fundamental Technological Research\vspace{-0.2cm} \\
\small Polish Academy of Sciences\vspace{-0.2cm}\\
\small \'Swi\c etokrzyska 21,  Warszawa  PL-00-049, Poland 
}

\date{\small \it (Submitted Sept. 1996  ------  Accepted July 1997) }

\begin{document}

\maketitle


\markboth{\centerline{\small\sl I.V. KANATCHIKOV}}{\hspace*{-12.5pt}
\centerline{\small\sl 
FIELD THEORY IN THE POLYMOMENTUM PHASE SPACE}} 


\newcommand{\beq}{\begin{equation}}
\newcommand{\eeq}{\end{equation}}
\newcommand{\beqa}{\begin{eqnarray}}
\newcommand{\eeqa}{\end{eqnarray}}
\newcommand{\nn}{\nonumber}

\newcommand{\half}{\frac{1}{2}}

\newcommand{\xt}{\tilde{X}}

\newcommand{\uind}[2]{^{#1_1 \, ... \, #1_{#2}} }
\newcommand{\lind}[2]{_{#1_1 \, ... \, #1_{#2}} }
\newcommand{\com}[2]{[#1,#2]_{-}} 
\newcommand{\acom}[2]{[#1,#2]_{+}} 
\newcommand{\compm}[2]{[#1,#2]_{\pm}} 
\newcommand{\lie}[1]{\pounds_{#1}}
\newcommand{\co}{\circ}
\newcommand{\sgn}[1]{(-)^{#1}}
\newcommand{\lbr}[2]{ [ \hspace*{-1.5pt} [ #1 , #2 ] \hspace*{-1.5pt} ] }
\newcommand{\lbrpm}[2]{ [ \hspace*{-1.5pt} [ #1 , #2 ] \hspace*{-1.5pt}
]_{\pm} }
\newcommand{\lbrp}[2]{ [ \hspace*{-1.5pt} [ #1 , #2 ] \hspace*{-1.5pt} ]_+ }
\newcommand{\lbrm}[2]{ [ \hspace*{-1.5pt} [ #1 , #2 ] \hspace*{-1.5pt} ]_- }
\newcommand{\pbr}[2]{ \{ \hspace*{-2.2pt} [ #1 , #2 ] \hspace*{-2.55pt} \} }
\newcommand{\we}{\wedge}
\newcommand{\dv}{d^V}
\newcommand{\nbrpq}[2]{\nbr{\xxi{#1}{1}}{\xxi{#2}{2}}}
\newcommand{\lieni}[2]{$\pounds$${}_{\stackrel{#1}{X}_{#2}}$  }
\newcommand{\rbox}[2]{\raisebox{#1}{#2}}
\newcommand{\xx}[1]{\raisebox{1pt}{$\stackrel{#1}{X}$}}
\newcommand{\xxi}[2]{\raisebox{1pt}{$\stackrel{#1}{X}$$_{#2}$}}
\newcommand{\ff}[1]{\raisebox{1pt}{$\stackrel{#1}{F}$}}
\newcommand{\dd}[1]{\raisebox{1pt}{$\stackrel{#1}{D}$}}
\newcommand{\nbr}[2]{{\bf[}#1 , #2{\bf ]}}
\newcommand{\der}{\partial}
\newcommand{\oo}{$\Omega$}
\newcommand{\Om}{\Omega}
\newcommand{\om}{\omega}
\newcommand{\eps}{\epsilon}
\newcommand{\si}{\sigma}

\newcommand{\inn}{\hspace*{2pt}\raisebox{-1pt}{\rule{6pt}{.5pt}\hspace*
{0pt}\rule{.5pt}{8pt}\hspace*{2pt}}}
\newcommand{\sro}{Schr\"{o}dinger\ }
\newcommand{\bm}{\boldmath}
\newcommand{\vol}{\widetilde{vol}}                                             \newcommand{\dvol}[1]{\der_{#1}\inn \vol}

\newcommand{\bd}{\mbox{\bf d}}
\newcommand{\bder}{\mbox{\bm $\der$}}
\newcommand{\bI}{\mbox{\bm $I$}}

\newcommand{\D}{\mbox{$\cal D$}}


\vspace*{-96mm}
\hbox to 6.2truein{
\footnotesize\it 
to appear in   
\hfil \hbox to 0 truecm{\hss 
\normalsize\rm July 1997 }\vspace*{-1mm}}
\hbox to 6.2truein{
\vspace*{-1mm}\footnotesize 
Rep. Math. Phys. 
\hfil 
} 
\hbox to 6.2truein{
\vspace*{-1mm}\footnotesize 
vol. {\bf 41}, No. 1 (1998)  \hfil 
\hbox to 0 truecm{ 
\hss \normalsize hep-th/9709229}
}
\vspace*{72mm}

\begin{flushright} 
\begin{minipage}{5.3in}
{\footnotesize
Canonical structure of classical field theory 
in $n$ dimensions is studied 
within the covariant polymomentum Hamiltonian  
formulation of De Donder--Weyl (DW).  
The bi-vertical $(n+1)$-form,  called  polysymplectic,
is put forward as a generalization of the symplectic form 
in mechanics.    
Although not given in intrinsic geometric terms 
differently than a certain coset 
it gives rise to an invariantly defined map between 
horizontal forms playing the role of dynamical variables 
and 
the so-called 
vertical multivectors 
generalizing Hamiltonian vector fields. 
The analogue of the Poisson bracket on forms is defined 
which leads to the structure of  
{$\Bbb Z$}-graded Lie algebra  
on 
the 
so-called Hamiltonian forms 
for which the map above exists.  
A generalized Poisson structure appears in the form of 
what we call   
a   
``higher-order'' and 
a   
right Gerstenhaber algebra.   
The equations of motion of forms are formulated in 
terms of the Poisson bracket with the 
DW Hamiltonian $n$-form 
$H\vol$ 
($\vol$ is the space-time volume form, $H$ 
is the DW Hamiltonian function) 
which is found to be related to the operation 
of
the 
total exterior differentiation 
of forms.   
A few applications and 
a relation to the standard Hamiltonian formalism in field theory 
are briefly discussed.\\
}   
\end{minipage}
\end{flushright} 

\section{\normalsize\bf Introduction}

The Hamiltonian formalism is based on the representation of 
the equations of motion in the first order form and the Legendre 
transform. The mathematical 
structures emerging from such a formulation of dynamics are known 
to be of fundamental importance in wide area of applications from 
the study of integrable systems to quantization. 

The generalization 
of  Hamiltonian formalism 
to field theory is well-known which is based on 
the functional derivative equations of first order in 
partial time derivative. 
This formulation 
requires an  explicit
singling out of the time, or evolution variable
and leads to the
idea of a field as a 
dynamical system with a continually infinite
number of degrees of freedom. 
The canonical quantization in field theory 
is known to be based on this approach.  
It can be given a covariant form a version of which is 
discussed for instance  in \cite{crnkovic}. 
Other discussions of the Hamiltonian formalism in field theory 
and further details can be found for example 
in  \cite{Abr+Marsden,chernoff,IZ,olver}.
Within this approach the geometrical constructions of 
classical mechanics can in principle be extended to field theory 
using the functional analytic framework of infinite-dimensional 
geometry, but the applicability of such constructions is often more 
restricted as, for example, 
known difficulties in geometric quantization  of field theory demonstrate. 
It is also not clear whether the framework of the canonical quantization based 
on the present Hamiltonian formalism; which requires the space-time to be, 
topologically, a direct product of  space and time, is adequate 
for  theories like General Relativity. 

However, another formulation of field equations in the 
form of  first order partial differential equations exists, 
the construction of which is also similar to the way the 
Hamiltonian formulation is constructed in mechanics, 
but which keeps the symmetry between space and 
time explicit.  
This formulation seems to be much less commonly known in theoretical 
physics in spite of the fact that its essential elements appeared more
than sixty years ago  in the papers  
by De Donder \cite{De Donder}, Carath\'{e}odory
\cite{Carath29}, Weyl \cite{Weyl35} and others 
on the multiple integral  variational calculus 
(see for example
\cite{Rund,Kastrup83,binz,giaq} 
for a review and further references).   
In this approach (see Sect. 2 for more details) 
the generalized coordinates are the field variables 
$y^a$ 
(not the field configurations $y^a(${\bf x}$ )$!) 
to which a set of $n$ momentum-like  variables, 
called {\em polymomenta},  
$p^i_a:=\partial L  
/
 \partial (\partial_iy^a)$,  is associated 
(here $i=1,...,n$ is the space-time index). 
Similar to mechanics, the covariant 
Legendre transform: $\der_i y^a \rightarrow p^i_a$, $L(y^a, \der_i y^a, x^i)
\rightarrow H_{DW}(y^a,p^i_a,x^i):= p^i_a\der_i y^a - L$ 
is performed,  
where the latter expression defines the 
covariant field theoretical analogue of Hamilton's canonical function 
which we will call the De Donder--Weyl (DW) Hamiltonian function.  
Unlike the Hamiltonian density in the 
standard (instantaneous) 
Hamiltonian formalism, 
which is the time component of the energy-momentum 
tensor, the DW Hamiltonian function is a scalar quantity 
a direct physical interpretation of which is not evident. 
What is interesting is that in terms of the variables above 
the Euler-Lagrange field equations 
take the form 
of the system of  first order partial differential equations 
(see eqs. (2.3) below) 
which naturally generalizes the Hamilton canonical equations of motion 
to field theory. This form of  field equations  
is entirely space-time symmetric and, in addition, 
is formulated in the finite dimensional covariant 
analogue of the phase space, 
the space of variables $y^a$ and $p^i_a$   
which we call here the {\em polymomentum} phase space 
(in our earlier papers the term ``DW phase space'' was used).     
By this means a field theory appears as a kind of generalized Hamiltonian 
system with many ``times'' the role of which is played both by space and 
time variables treated in a completely symmetric manner. 
In doing so note that the DW Hamiltonian function does not generate 
a time evolution of a field from a given Cauchy data, as the standard 
Hamiltonian does, but rather controls a space-time variation, or 
development,  of a field. 

Because of the fundamental features which the  formulation 
outlined above 
shares with the Hamiltonian formulation of mechanics 
we  refer to it as  the DW Hamiltonian formulation.  
Surprisingly enough, 
basic elements of a possible canonical formalism based on this formulation, 
and even the existence of some of these,  still are rather poorly understood. 
In fact, at present it is not even evident whether or not 
the DW formulation 
of field equations does indeed provide us with a starting point for 
a certain canonical formalism in field theory, with an appropriate 
covariant analogue of the symplectic or Poisson structure, Poisson brackets  
and the related geometrical constructions, as well as the starting point 
for a  quantization.

Recall that there exists also 
the Hamilton-Jacobi theory inherently related to
the DW Hamiltonian formulation of field equations 
(see e.g. \cite{Rund,Kastrup83} for a review).   
This theory is formulated in terms of the certain covariant
partial differential equation for $n$ Hamilton-Jacobi 
functions  
$S^i(y^a, x^i)$: $\der_i S^i = H_{DW}(y^a,
p^i_a:= \der_a S^i, x^i) $ 
which obviously reduces to the 
familiar Hamilton-Jacobi  equation in mechanics when $n=1$. 
However, while the 
connection between the Hamilton-Jacobi  equation and 
the Schr\"odinger equation in quantum mechanics is a well 
known fact 
which underlies both the quasi-classical approximation 
and the De Broglie-Bohm interpretation of quantum mechanics,    
a similar possible connection between 
the DW Hamilton-Jacobi equation and 
quantum field theory 
still remains completely unexplored.  
In this context it is 
worth remembering that  
historically the arguments based 
on 
 the 
Hamilton-Jacobi theory led Schr\"odinger 
to his famous equation, 
and Wheeler to the Wheeler--DeWitt equation in quantum gravity. 
It is quite  natural to ask, therefore, 
whether the DW Hamilton-Jacobi equation also
may help us to reveal a 
certain new aspect of quantum field theory.

It should be noted that the DW canonical theory is, in fact, 
only the simplest representative of the whole variety of  
covariant canonical formulations of field theories 
in a polymomentum 
analogue of the phase space 
which are based on different choices of the
"Lepagean equivalents" of the so-called Poincar\'e-Cartan form
(see e.g. \cite{Kastrup83,Gotay ext,Gotay multi1,rund85}) 
and different definitions of polymomenta.
The famous canonical theory of Carath\'eodory
\cite{Carath29,Rund,Kastrup83,rund85}
is another interesting example of such a formulation.  
A discussion of the corresponding  
more general ``polymomentum canonical theories'' 
for fields,  as we suggest to name them, 
is left beyond the scope of the present  paper.

Despite all of the attractive features of 
polymomentum  canonical theories, 
such as  finite dimensionality and a manifest covariance, 
which seem  especially relevant in the context 
of the canonical analysis and quantization 
of general relativity and string theory 
(note, that no  restriction to 
the globally hyperbolic  
underlying space-time manifold is in principle implied here),
there are a surprisingly small number of their applications
in the literature.  The following papers 
contain  some applications to 
classical field theory 
\cite{Kij+Tul,Rieth,Grigore sour},
gauge fields \cite{Kij84,Sardan}, 
classical bosonic string 
\cite{Kastr+Rinke,Nambu80,Beig,Kanatch,Grigore ngoto},
general relativity \cite{Francav,Horava,napoli},  
and integrable systems \cite{dickey,gotay}. 
Several interesting  examples are also considered in 
the book \cite{Gotay ea}. 

For possible applications in field theory 
the understanding of the interrelations  between 
the polymomentum  
canonical theories and the standard instantaneous  
Hamiltonian formalism is important. 
A recent discussion of this body of questions 
can be  found in the papers by Gotay  
\cite{Gotay multi2} 
(see also the preprint of the book \cite{Gotay ea}  
and the paper by \'Sniatycki  \cite{sniat}). 
Let us   note also that the 
standard  functional Hamiltonian form of  
field equations 
can in principle be derived directly from 
the DW covariant Hamiltonian equations \cite{kij-private}.   

Among   
the questions concerning the applications of 
the polymomentum canonical theories 
one of the most interesting ones 
is 
whether it is possible, or has any sense,  
to develop a field quantization
starting from
the DW Hamiltonian formulation 
or a more general
polymomentum  canonical theory.  
Indeed, 
it might be questioned if it so necessary to
at first split  the space-time in order to obtain the
Hamiltonian formulation, 
and then to quantize a field according to the 
standard prescriptions of quantum theory 
and to prove finally 
the procedure to
be consistent with the relativistic symmetries. 
Or, perhaps,   
it is possible to develop an inherently covariant 
field quantization based on 
the polymomentum 
covariant Hamiltonian framework,  
without altering the space-time symmetry, 
and then  to obtain 
the results referring to a particular reference frame, 
if necessary, from the manifestly
covariant formulation of quantum field dynamics.
Another related question is whether
a sort of
quasi-classical transition from 
some 
formulation of
quantum field theory to 
the Hamilton-Jacobi equations 
corresponding to
various polymomentum  canonical formulations 
of classical fields 
(see \cite{Rund,Kastrup83}) exists. 
Obviously, to approach these questions one
has to gain a deeper insight into those geometric and algebraic
structures of  classical polymomentum canonical theories  
whose analogues in mechanics   
form a classical basis of the quantization procedures. 
This paper may be considered 
as a step in this direction.

Recall that the problem of field quantization based
on the DW Hamiltonian formalism was
briefly discussed in the middle of the thirties 
by Born \cite{Born34} and Weyl
\cite{Weyl34}.  Then, in the early seventies 
 considerable progress was made in  
understanding the differential geometric structures 
underlying  the De Donder--Weyl theory 
\cite{Garcia,Gold+Stern,Kij ea,Gaw} 
(see also  earlier papers by Dedecker  \cite{Dedecker} 
who studied more general canonical 
theories,  and the recent paper by Gotay
\cite{Gotay ext} for a  subsequent development) 
which led to the so-called 
multisymplectic formalism \cite{Kij ea}. 
However, 
the known
attempts \cite{Herm lie,Garcia,Kij ea,Gaw}
to approach a field quantization  from this 
viewpoint  essentially  concentrated on 
only establishing links with the 
conventional formulation  based on the
instantaneous Hamiltonian  formalism and 
have not led  to any new formulation.    
More recently, an  attempt to 
construct a quantization of  field theory 
based entirely on the polymomentum    
canonical framework  was reported by 
G\"{u}nther in \cite{Guenther87b} who 
used his own geometrical version of 
De Donder-Weyl  
theory -- the so-called "polysymplectic Hamiltonian formalism" 
\cite{Guenther87a}.  
Unfortunately, the  
ideas of his brief report \cite{Guenther87b} were not developed 
to the point where a comparison 
with results of  the conventional quantum field theory  
would be possible.  
A few other related discussions  may be also found in 
recent papers  \cite{sard94,good95,nav95}.

The main obstacle  to the development  of
field quantization based on 
a polymomentum  Hamiltonian formulation 
seems to be the lack of
an 
appropriate
generalization  of the Poisson bracket. 
Within the 
multisymplectic formalism
\cite{Gotay multi1,Gotay multi2,Gotay ea,Kij+Tul,Kij ea} 
which is closely related to the DW canonical theory 
a  Poisson bracket was proposed in   
\cite {Gold+Stern,Kij ea,Gaw}
which is defined on forms of degree $(n-1)$ corresponding 
to observables in field theory 
(see also \cite{Herm lie} and 
the recent paper \cite{crampin}). 
However,  
the related construction proved to be too restrictive to reproduce the
algebra of observables, or currents,  
in the theories of a   sufficiently general type 
\cite{Gold+Stern}, 
and it was not 
appropriate for representing the DW Hamiltonian field
equations in Poisson bracket formulation. 
Moreover, the Jacobi identity for
this bracket was found to be fulfilled modulo the exact terms only
(see e.g. \cite{Gold+Stern,Gotay ea}), although this fact 
may seem interesting from the point of view of homotopy Lie algebras.  
Other approaches,  by Good 
\cite{Good}, Edelen \cite{Edelen} and 
G\"{u}nther \cite{Guenther87a},
enable 
in principle writing of DW Hamiltonian field equations in a certain 
bracket formulation, 
but the algebraic properties of the brackets introduced 
by these authors 
and, therefore, their usefulness for quantization are rather obscure.
Another relevant discussion of a covariant Poisson bracket in field theory 
may be also found in 
\cite{marsden ea}.    

\medskip 

The purpose of the present study is to develop those elements of
the 
DW canonical theory whose analogues in the Hamiltonian formalism
of mechanics are important   
for  canonical or geometric quantization.   
These in particular include the symplectic form, the Poisson bracket,  
the notion of canonically  conjugate variables, and the representation of 
the equations of motion in Poisson bracket  formulation.
There is probably no more reliable basis  for 
attacking this problem 
than to start 
from the most fundamental object of any canonical theory -- 
the Poincar\'{e}-Cartan (PC) $n$-form -- 
and to
try to develop the subsequent elements of the formalism
by searching
for the proper generalizations 
to the DW formulation of field theory 
of the corresponding  elements of the
canonical formalism of mechanics 
(as they are presented, for example, 
in classical texts \cite{Abr+Marsden} and \cite{Arnold}).

\medskip

The structure of the paper is as follows. 
In Sect. 2 we demonstrate how the DW Hamiltonian  
field equations readily follow from
the PC form. This consideration indicates 
a suitable generalization to field theory 
of the notion of the canonical Hamiltonian vector field.
This is the multivector field  
of degree $n$  whose integral $n$-surfaces in the 
extended polymomentum 
phase space of the DW theory represent the extremals of the 
variational problem describing a field, 
that is   the solutions of field equations. 
We point out that the vertical components of this multivector 
field essentially contain all information about the equations of motion. 

The latter  fact leads us to the notion of 
the {\em polysymplectic} form in Sect. 3. 
This is a  closed "bi-vertical" 
form of degree $(n+1)$, eq. (10), 
which is proposed as a generalization of the symplectic 
form to the DW Hamiltonian formulation of field theory. 
The function  of the polysymplectic form is that it 
provides us with the map between vertical multivector fields  
(generalizing  Hamiltonian vector fields in mechanics) 
and the so-called Hamiltonian horizontal forms 
(which play the role of dynamical variables).  
We construct the bracket operations on 
 Hamiltonian multivector fields 
and Hamiltonian forms 
and show that they endow  the corresponding spaces 
with the structure of 
\mbox{{$\Bbb Z$}}--graded Lie algebra. 
We next discuss two generalizations of the derivation property of the 
usual Poisson bracket 
to the 
bracket operation 
on differential forms (which is called the graded Poisson bracket) 
 introduced here. 
This leads to what we call respectively  
a higher-order and a right Gerstenhaber algebra 
as generalizations of a Poisson algebra 
to the DW Hamiltonian formulation of field theory.  


The graded Poisson bracket of forms  
is used in Sect. 4 for  obtaining the equations 
of motion of Hamiltonian forms of degree $(n-1)$ 
in Poisson bracket formulation. 
As a by-product, 
the proper generalization of the notions of an integral
of the motion and of canonically conjugate variables 
to the DW formulation is given. 

In Sect. 5  the bracket representation of
the equations of 
motion of arbitrary forms is considered. 
For this purpose the bracket with  forms of degree $n$ 
is defined 
which  requires the enlargement of 
the space of Hamiltonian multivector fields by  
vertical-vector valued 
one-forms.  
We show that the equations of motion may be 
written in terms of the bracket with the DW $n$-form $H \vol$  
and shortly discuss the problems 
related to the algebraic closure of the enlargement above. 

In Sect. 6,  a few examples of  applications of our 
graded Poisson bracket on forms 
to 
the system of interacting scalar fields, 
electrodynamics and 
the Nambu-Goto string are considered.  
A general discussion of our results,  
which also includes observations concerning 
a relation of the graded Poisson bracket on forms 
to the standard Poisson bracket of functionals,
 is presented in Sect. 7. 
In App. 1 some details of the calculation 
related to the higher-order graded Leibniz rule can be found. 


\section{\normalsize\bf De Donder--Weyl 
Hamiltonian field equations 
and the 
Poincar{\'e}--Cartan form }

Let us consider a field theory given by the first order 
variational problem
\begin{equation}
\delta \int L(y^a, \partial_iy^a, x^i)\widetilde{vol}=0, 
\end{equation}
where $\{y^a\}, 1\leq a\leq m$ are field variables, 
$\{x^i\}, 1\leq i\leq n$ are space-time variables, 
and 
\[\widetilde{vol}:=dx^i\wedge...\wedge dx^n , \] 
is the volume form on the space-time manifold. 
In order to simplify formulae, 
in the definition of the 
$n$-volume form $\vol$ 
and 
the following discussion 
it is implied that the coordinates on 
the $x$-space are chosen so that the metric determinant $|g|=1$.

The standard way of studying the solutions of the variational problem 
is solving the Euler-Lagrange equations. 
A more geometrical approach 
is formulated in terms of the so-called Poincar\'e--Cartan form which is 
usually written in terms of the Lagrangian coordinates  
$(y^a, \partial_iy^a, x^i)$. 
For our  purposes, however, 
we introduce a new set of Hamiltonian like variables:
\[p^i_a:=\frac{\partial L}{\partial(\partial_iy^a)} \] 
and
\begin{equation}
H_{DW}(y^a,p^i_a, x^i):=p^i_a\partial_iy^a-L
\end{equation}
which are  defined in a completely space-time symmetric
manner. 
The set of variables above is central to the so-called 
De Donder--Weyl (DW) 
canonical theory for fields which  appeared
for the first time     
in the papers on the calculus of variations 
in the middle of the 1930s
\cite{De Donder,Weyl35}. 
We shall call these variables  the {\em polymomenta} 
and 
the {\em DW Hamiltonian function} respectively. It is a
rather straightforward  calculation to demonstrate that 
in terms of the variables above the Euler-Lagrange 
equations may be written in the following first order form
(see e.g. \cite{Rund,Kastrup83,binz,Gotay multi1,Gotay ea}) 
\beq 
\partial_ip^i_a=-\frac{\partial H}{\partial y^a}, 
\quad 
\partial_iy^a=\frac{\partial H}{\partial p^i_a}.
\eeq 
This is the canonical form 
of field equations within 
the DW theory. 
In view of the evident similarity of eqs. (2.3) 
to Hamilton's equations in mechanics 
(whose multidimensional, or rather ``multi-time''  
generalization they provide us with)
we will call them the   {\em DW Hamiltonian field equations.}
Obviously, this formulation of field equations is explicitly space-time 
symmetric, and, unlike the instantaneous approach, it involves 
the {\em finite dimensional} analogue 
of the phase space, the $(m+mn+n)$--dimensional space of variables 
\[
z^M:=(y^a, p^j_b, x^i), \quad 1\leq M \leq m+mn+n 
\]
which we call the {\em extended polymomentum  phase space.}

The present paper is devoted to the investigation of the mathematical 
structures underlying the above formulation 
of field equations.
 The question of particular interest for us is whether  
a generalization of the Poisson brackets to the DW Hamiltonian formulation
exists and whether the DW Hamiltonian field equations can be written in terms 
of these brackets. In mechanics the answers to these questions are well 
known and they lead to the geometrical description of mechanics in 
terms of the symplectic or the Poisson structure 
(see e.g. \cite{Abr+Marsden,olver,Arnold}). 
This description is known to have its roots
in the geometrical formulation of the
one dimensional calculus of variations in terms of the 
Poincar\'e--Cartan (PC) form.  
In a certain sense, all the elements of 
the canonical formalism in mechanics can be obtained step by 
step proceeding directly from the PC form. 
In field theory 
considered from the point of view of the 
DW canonical theory 
the analogues of the corresponding 
constructions are  either unknown or rather poorly understood 
while the notion of the PC form 
is  well defined, at least for first order theories.  
In this paper we  try to recover 
the structures underlying the DW formulation of field theory 
by proceeding from the corresponding PC form and 
developing the elements of the canonical formalism 
based on the analogy with the corresponding 
constructions in mechanics. 

As the first step of realization of this idea 
one has to find an appropriate description of
the DW Hamiltonian field equations in terms of the PC form. 
In mechanics the equations of motion appear as the equations 
for the integral curves of the
total canonical Hamiltonian vector field 
which annihilates the exterior differential of
the PC one-form.  
A field theoretical generalization of this
fact is presented below.

The PC form in terms of the DW Hamiltonian variables is 
known to be given by 
(see e.g. Refs. \cite{Kastrup83,Gotay ea,Gold+Stern}) 
\[
\Theta_{DW}=p^i_a\wedge dy^a\wedge\partial_i\inn\vol - H_{DW}\vol . 
\]
Its exterior differential, the canonical $(n+1)$-form,  
is 
\begin{equation}
\Omega_{DW}=dp^i_a\wedge dy^a\wedge\partial_i\inn\widetilde{vol}-dH_{DW}
\wedge\widetilde{vol}. 
\end{equation}
The symbol \inn\  denotes the interior
product of a (multi)vector 
to its left and a form to its right.  
In the following, we will omit the subscript DW in $H_{DW}$, 
but the scalar quantity 
to be denoted as $H$ 
and termed the (DW) Hamiltonian function 
should not be 
confused with the usual Hamiltonian or its density 
which will not appear in this paper.    

The canonical $(n+1)$-form $\Omega_{DW}$ contains 
all the information about 
a field's  dynamics, because it is constructed immediately from the 
Lagrangian density. 
In particular, the DW Hamiltonian form 
of field equations 
can be  derived directly from $\Omega_{DW}$. 
Recall that from the calculus of variations 
it is known that 
the extremals of the variational 
problem are 
isotropic subspaces of the PC form
spanned
by the (vertical) vector fields annihilating $\Omega_{DW}$ 
(cf. e.g.
\cite{Kastrup83,Gotay multi1,Gotay ea,Kij+Tul,Gold+Stern,Herm lie,crampin}).
For our purposes another formulation of this fact proved to be useful. 
Namely,  the solutions of the variational problem (2.1) 
may be viewed as 
$n$-dimensional 
	surfaces 
in the extended polymomentum  phase space.  
Let us describe these 
	surfaces  
by the multivector field of degree $n$ 
(or $n$-vector, for brevity) denoted \xx{n}  
\begin{equation}
\xx{n}:=   
\frac{1}{n!}
\stackrel{n}{X}{}^{M_1...M_n}(z)\,
 \der\lind{M}{n}
\end{equation}
which represents their tangent $n$-planes.    
Here the notation 
\[\der\lind{M}{n}:=\partial_{M_1}
\wedge...\wedge\partial_{M_n}\]
is introduced.
The $n$-vector field 
$\stackrel{n}{X}$ naturally generalizes 
the velocity
field of the canonical Hamiltonian flow in classical mechanics 
to the multidimensional, $n>1$,  
case of field theory.

The condition on
$\stackrel{n}{X}$ to determine  
classical extremals 
is that the form
$\Omega$$_{DW}$  should vanish on \xx{n},  
that is 
\begin{equation}
\mbox{\xx{n}\inn\ $\Omega_{DW}=0, $ } 
\end{equation}
or in terms of the components of $\xx{n}$
\beqa
\xx{n}\inn\ \Omega&=&
\frac{1}{n!}
[X\uind{i}{n}\der \lind{i}{n} 
\nn \\ 
&+& 
X^{a}{}\uind{i}{n-1}\der_a{}\lind{i}{n-1}    +
X^i_a{}\uind{i}{n-1}\der^a_i{}\lind{i}{n-1}  \nn \\ 
&+& 
X^a{}^i_b{}\uind{i}{n-2} \der_a \we \der_i^b \we \der\lind{i}{n-2}
+ \quad ...
\quad 
]
\inn\ \Omega_{DW} = 0.  \nn  
\eeqa 
This gives rise to the following expressions 
\beqa
n X^a{}\uind{i}{n-1} \eps_i{}\lind{i}{n-1}&=& 
\der_i^a H X\uind{i}{n}\eps \lind{i}{n}      ,   \nn \\
n X^i_a{}\uind{i}{n-1} \eps_i{}\lind{i}{n-1}&=& 
\der_a H X\uind{i}{n}\eps \lind{i}{n}        ,     \\
(-1)^n (n-1) X^a{}^i_b{}\uind{i}{n-2}\delta^b_a 
\eps\lind{i}{n-2}{}_{ij}&=& 
\der_v H X^v{}\uind{i}{n-1}\eps \lind{i}{n-1}{}_j  .  \nn 
\eeqa
Further, the  
 integral $n$-surfaces
of the multivector field $\xx{n}$ 
can be  described by the differential equation 
written in  terms of the Jacobian 
\begin{equation}
\mbox{$\stackrel{n}{X}$$^{M_1...M_n}(z) =
{\cal N}$}  \frac{\partial(z^{M_1},...,z^{M_n})}
{\partial(x^1,...,x^n)},
\end{equation}
where $\cal N$ depends on the choice of 
the parameterization of 
an $n$--surface in the extended polymomentum phase space 
which represents a solution of field equations.   
Obviously, the equation above is a natural generalization 
of the o.d.e. for the integral curves of  
a 
vector field. 
Substituting the expression 
for  the components of $\xx{n}$, 
eq. (2.7), into (2.8) 
we obtain the DW Hamiltonian field equations, eqs. (2.3), 
from the ``vertical'' components
$X{}^{ai_1 ... i_{n-1}}$ and 
$X{}^{ii_1...i_{n-1}}_{a}$ of $\xx{n}$.
The ``bi-vertical'' components,   
\xx{n}$^{aii_1 ... i_{n-2}}_{\cdot a}$, lead to a consequence  
of the DW field equations. 
Thus, the information about the classical dynamics of fields is
essentially encoded in the "vertical" 
components of the multivector field 
annihilating the 
canonical $(n+1)$-form (2.4).   
This  observation  underlies our  construction in the 
following section.  
Note that the issue of the integrability of the multivector field 
$\xx{n}$ essentially does not arise here 
because the condition (2.6) specifies in fact only a small fraction  
of the components of $\xx{n}$ (cf. eq. (2.7)),   
so that the remaining components are only restricted by the 
consistency of  eq. (2.8).    


\section{\normalsize\bf Polysymplectic form, Hamiltonian multivector fields 
and forms,       and graded Poisson brackets}

In this section a generalization
of the basic structures of classical Hamiltonian mechanics, 
such as the symplectic form, 
Hamiltonian vector fields and functions, 
and the Poisson bracket,   
to the DW Hamiltonian formulation of field theory is presented. 

Our starting point is the observation  that the well-known formula 
$X_H \inn\ \omega = d H$ 
relating the canonical Hamiltonian 
vector field $X_H$ 
to Hamilton's function $H$ 
by means of the map 
given by the symplectic form $\omega$ 
has its natural 
counterpart, eq. (3.2),  in the DW theory. 


For this purpose let us note that the 
extended polymomentum phase space can be viewed as 
a bundle $\pi: {\cal Z}\rightarrow M$ over the space-time manifold $M$.   
The local coordinates on $ {\cal Z}$ are $\{ z^M \}$ introduced above. 
The fiber coordinates 
\[
z^v=(y^a,p^i_a), \quad v=1,...,m, m+1,...,m+mn 
\]
will be referred to as {\em vertical}, 
as well as the corresponding subspace and the objects related to it. 
Correspondingly, the space-time 
coordinates $x^i$ are  to be referred to as {\em horizontal}. 
\newcommand{\newa}{
In order to be able to distinguish explicitly the objects referring to 
the vertical and the horizontal subspaces  
To proceed we make a simplifying assumption that the {\em extended\/} 
polymomentum phase space is a trivial bundle over the space-time. 
It allows us to distinguish explicitly the objects referring to 
the vertical and the horizontal subspaces, which we find necessary at 
this stage. 
It also allows us to avoid additional problems with the 
intrinsic definition of the objects we are using. This simplification 
seems to be reasonable as here our 
purpose is rather to find out the way in which the constructions of the 
canonical formalism in mechanics can be extended to the framework of the 
De Donder--Weyl theory, and already in the case of the trivial bundle 
structure there are many  questions to be answered.
} 

Now, let us  introduce several objects 
which will appear in the subsequent  discussion. 
The definitions will be  given  
in local coordinate terms 
used throughout the present paper. 
Although we will also make some remarks  
concerning the invariant meaning of 
the introduced objects, 
the issue of the intrinsic geometric formulation 
of our  approach is left beyond the scope of this 
paper.  

 (i) A {\em vertical} vector  $X^V$ is 
an element of the vertical tangent bundle of ${\cal Z}$,   
$V$$T$${\cal Z}:= \ker T \pi$. In local coordinates 
\[ X^V:= X^a  \der_a + X^i_a  \der^a_i .\] 

(ii) A form of degree $p$, $\ff{p}$, 
is called  {\em horizontal\/} 
if in local coordinates it takes the form 
\[\ff{p}:=
\frac{1}{p!}
F_{i_1 ... i_p}(z) \, dx^{i_1}\wedge ...\wedge dx^{i_p}.
 \]  
Intrinsically this means that a contraction of a horizontal form  with 
a vertical vector always vanishes. 
More generally, a form of degree $p$ is called $(p-q)$-horizontal if 
its contraction with any $(q+1)$ vertical vectors vanishes. This means 
that it has $(p-q)$ or more horizontal components. The space of 
$(p-q)$-horizontal  $p$-forms is to be denoted $\bigwedge^p_q$. 

(iii) A multivector field of degree $p$ is called {\em vertical} 
and denoted  $\xx{p}{}^V$ 
if in a local coordinate system it can be written in the form 
\[\xx{p}{}^{V}:=
\frac{1}{p!}
\xx{p}{}^{vi_1...i_{p-1}}(z) \, 
\der_{v}\wedge\der_{i_1}\wedge...{}\wedge\der_{i_{p-1}} .\]
This definition is not invariant, 
as in a different coordinate system higher vertical components 
may appear. The intrinsic definition is that the 
inner product of  the vertical $p$-multivector with any 
horizontal $p$-form vanishes. This means that $\xx{p}{}^{V}$  has 
{\em at least\/}  one vertical component.   

\medskip
(iv) For an arbitrary $p$ form 
$\Phi = \frac{1}{p!}\Phi \uind{M}{p} dz^{M_1}\we ... \we dz^{M_p}$, 
where $\{dz^M\}:=\{dz^v,dx^i\}:=\{dy^a,dp^i_a,dx^i\}$ 
is a holonomic basis of $T^*{\cal Z}$,  
the operation  of the 
{\em vertical exterior differentiation},  $\dv$, 
in local coordinates is given by 
\[
\dv \Phi = \frac{1}{p!}\der_v \Phi\uind{M}{p} 
dz^v\we dz^{M_1}\we ... \we dz^{M_p}.
\]
In particular, 
\[d^{V}H = \der_aH dy^a + \der^a_iH dp^i_a. 
 \]
As under the coordinate transformation 
$dz^v$ transform as 
$dz'{}^v=\frac{\der z'{}^v}{\der z^w}dz^w + 
\frac{\der z'{}^v}{\der x^i }dx^i$ 
the expressions above are not invariant.   
However, if $\Phi \in \bigwedge^p_q$  
and $d$ denotes the exterior differential 
on the extended polymomentum phase space ${\cal Z}$,   
the expression $\dv \Phi$ 
can be understood as 
the coset 
$\dv \Phi:= [d \Phi \quad mod \bigwedge^{p+1}_q$], 
so that  the highest horizontal part is factored off.   
\newcommand{\newb}{
$\xx{p}{}^V$, called 
{\em vertical multivectors} of degree $p$, 
which are defined to have the following 
representation in a local coordinate system:
\[\xx{p}{}^{V}:=
\frac{1}{(p-1)!}
\xx{p}{}^{vi_1...i_{p-1}}(z)
\der_{v}\wedge\der_{i_1}\wedge...{}\wedge\der_{i_{p-1}} .\]
We also need to define the operation of the 
{\em vertical exterior 
differentiation} $d^V$ 
which for  any form $F$ is given by 
\[d^V F := dz^v \wedge \der_v F,
      \] 
where $\der_v$ denotes the
partial differentiation of the components of $F$ with respect
to the vertical variables. In particular,
\[d^{V}H = \der_aH dy^a + \der^a_iH dp^i_a. 
 \]
In the following 
the notion of a {\em horizontal $p$-form}
$p$-form will also be used  which in local coordinates 
takes the form 
\[\ff{p}:=
\frac{1}{p!}
F_{i_1 ... i_p}(z) dx^{i_1}\wedge ...\wedge dx^{i_p}.
 \]
}

Further, 
let us separate out the non-horizontal part of the PC-form 
in Sect. 2: 
\[
 \Theta^V := p^i_a dy^a \wedge \dvol{i}. 
\] 
This can be done in the simplified, yet nonetheless interesting, 
case of the trivial 
bundle structure of the extended polymomentum phase space.  
In a more general situation the expression above can be 
understood as the coset:   
\mbox{$\Theta^V:= [\Theta \quad mod \bigwedge^n_0]$}.

By taking the vertical exterior differential of $\Theta^V$   
we obtain 
$\Omega^V := d^V\Theta^V,$
which  in local coordinates can be written as follows   
\beq
\Omega^{V} := -dy^a\we dp^i_a \we \der_{i} \inn \widetilde{vol} .
\eeq
This object is not a form unless 
the extended polymomentum phase space 
is a trivial bundle over the space-time. 
In a more general case  it can be treated as 
the  coset  
$\Omega^{V} := [\Omega_{DW} \quad mod \bigwedge^n_1]$. 

\medskip

Now, 
the geometrical condition 
(2.6)
which leads  to the DW field equations can be 
written  
in the form 
\begin{equation}
\xx{n}{}^V\inn\ \Omega^V = (-1)^n\, d^{V}H , 
\end{equation} 
if the parameterization in 
(2.8) is chosen in 
such a way that 
\[
\frac{1}{n!}
\xx{n}{}\uind{i}{n}\der\lind{i}{n}\inn\ \vol =1. \]
This observation justifies a significance of the objects 
introduced above and inspires the construction below.

We shall call the object  $\Omega^V$ in (3.1)  
{\em the polysymplectic form} 
adopting the term earlier introduced by Guenther \cite{Guenther87a}
for a similar object -- the horizontal-vector valued vertical two-form
$dy^a \we dp^i_a \otimes \der_i$ -- 
in the context of his polysymplectic Hamiltonian formalism. 
The polysymplectic form defined here 
has a potential advantage that 
it is related to the multidimensional PC form of the calculus 
of variations    
in exactly the same way as 
the symplectic form in mechanics is related to
the one dimensional PC form.  
In what follows, the 
 polysymplectic form 
$\Omega^V$ 
will be denoted simply as $\Omega$, 
and 
the superscripts $^V$ 
labeling the vertical 
multivectors will also be dropped, 
since all the multivectors in the subsequent discussion 
are vertical. 
Note also that the subsequent relations involving 
$\dv$ and $\Omega$ have to be understood as 
the relations between the corresponding equivalence classes 
(modulo terms of the highest horizontal degree).    

\newcommand{\oldb}{
In what follows, the 
$(n+1)$-form $\Omega^V$ 
is denoted simply as $\Omega$. We 
call it the {\em polysymplectic} form
adopting the term introduced by Guenther \cite{Guenther87a}
for a similar object -- the horizontal-vector valued vertical two-form
$dy^a \we dp^i_a \otimes \der_i$ -- 
in the context of his polysymplectic 
Hamiltonian formalism. 
Note that the polysymplectic form introduced above in
eq. (3.1) 
is related to the PC form  
in exactly the same way as 
the symplectic form in mechanics is related to
the PC form of the one-dimensional variational calculus. 
}

\newcommand{\oldbb}{
Before proceeding further let us 
note  
that the objects introduced 
above are given in local coordinate terms. The way they are defined 
is suitable if the extended polymomentum phase  space is a globally 
trivial bundle over the space-time. This simplification is valid for 
scalar fields but not for arbitrary tensor fields. Here we use it 
as it allows us to  explicitly distinguish objects 
referring to 
the vertical and the horizontal subspaces, which we found necessary 
here. 
It also allows us to avoid supplementary problems with an  
intrinsic definition of the objects we are using. 
Because already in the case of the trivial bundle structure it is 
not clear how the constructions 
of the canonical formalism in mechanics could be generalized 
to the framework of the De Donder--Weyl theory, the simplification 
under discussion seems to be reasonable at this stage. 
}   

\newcommand{\newc}{
To proceed we make a simplifying assumption that the {\em extended\/} 
polymomentum phase space is a trivial bundle over the space-time. 
It allows us to distinguish explicitly the objects referring to 
the vertical and the horizontal subspaces, which we find necessary at 
this stage. 
It also allows us to avoid additional problems with the 
intrinsic definition of the objects we are using. This simplification 
seems to be reasonable as here our 
purpose is rather to find out the way in which the constructions of the 
canonical formalism in mechanics can be extended to the framework of the 
De Donder--Weyl theory, and already in the case of the trivial bundle 
structure there are many  questions to be answered.
}

\medskip

Let us now recall (see e.g. \cite{Abr+Marsden,Arnold} for details)  
that the structures of classical Hamiltonian mechanics are 
incorporated essentially
in a single statement, 
that Lie derivative of a symplectic form $\omega$  
with respect to the vertical vector fields $X$ 
which generate the infinitesimal canonical transformations vanishes: 
$\pounds$$_X \omega =0$. 
This is the well-known canonical symmetry of mechanics. 
Since $\omega$ is closed, 
it implies locally that $X_F \inn\
\omega =dF$ for some function $F$ of the phase space variables. 
If the latter 
equality holds globally, the vector field $X_F$ is said to be
a (globally) Hamiltonian vector field associated with the Hamiltonian
function $F$. When $F$ is taken to be 
the canonical Hamiltonian function $H$, 
the equations for the integral curves of $X_H$, the canonical
Hamiltonian vector field, 
reproduce Hamilton's canonical equations of motion.  
Further, the canonical symmetry and the map above allow to construct 
appropriate brackets on vector fields and functions.  
Our intention here is to find  an analogue of this 
scheme for the DW Hamiltonian formulation in field theory.

{}From the above considerations, it can already be  
concluded that
the vertical $n$-vector field \xx{n} associated with the DW
Hamiltonian function according to  eq. (3.2) 
is similar to the canonical
Hamiltonian vector field in mechanics, 
and that the polysymplectic form $\Omega$ 
is analogous to  
the symplectic 2-form.  
To pursue this parallel further, 
let us introduce  the operation of 
generalized Lie derivative  
with respect to  
a  vertical multivector field 
of degree $p$, $\xx{p}$,  
and postulate, as a fundamental symmetry principle 
extending the canonical symmetry of mechanics, that
\beq
\mbox{$\pounds$\rbox{1pt}{$_{\stackrel{n}{X}}$} $\Omega =0$} .
\eeq
A generalized Lie derivative of any form $\Phi$ 
with respect to the vertical multivector field 
\rbox{1pt}{$\stackrel{p}{X}$}  
of degree $p$ 
is given  by 
\beq
\mbox{$\pounds$\rbox{1pt}{$_{\stackrel{p}{X}}$} $\Phi :=$
\rbox{1pt}{$\stackrel{p}{X}$}\inn\ 
$d^V\Phi - 
(-1)^p \, d^V$(\rbox{1pt}{$\stackrel{p}{X}$}\inn\ $\Phi)$}
\eeq
which  is the simplest generalization of the Cartan formula relating
the Lie derivative of a form along the vector field to the exterior
derivative and the inner product with a vector.  
However,  unlike the $p=1$ case 
the operation $\pounds$$_{\stackrel{p}{X}}$ does not preserve 
the degree of a form it acts on. 
Instead, it maps $q$-forms to $(q-p+1)$-forms.
Note, that a similar definition of  
generalized Lie derivative 
with respect to a multivector field was considered earlier 
in   \cite{tulcz}.  

\medskip

   Since $\Omega$ is closed with respect to the vertical exterior
differential,  from 
(3.3) and (3.4) it follows that
\beq
\mbox{$d^{V}$ (\rbox{1pt}{$\stackrel{n}{X}$} \inn\ $\Omega)=0,$}
\eeq
{\rm so that locally we can write}
\begin{equation}
\mbox{\rbox{1pt}{$\stackrel{n}{X}$}\inn\ $\Omega=d^{V}$\rbox{1pt}
{$\stackrel{0}{F}$} , }
\end{equation}
for some $0$-form $\stackrel{0}{F}$  which depends on the 
polymomentum phase space  variables $z^M$. 
By analogy with mechanics,  
we can call \xx{n}  
the 
(globally) Hamiltonian $n$-vector field
associated with the Hamiltonian $0$-form \ff{0},
if such a form exists globally.
Similarly, the multivector field \xx{n}
satisfying eq. (3.3) or (3.5) can be  called locally Hamiltonian. 
From eqs. (3.2) and (3.6) 
we see that our symmetry postulate in eq. (3.3), 
together with the
definition 
of generalized Lie derivative in eq. (3.4), 
is consistent with the DW Hamiltonian field equations when 
the $0$-form \ff{0} in eq. (3.6) is taken to be 
the DW Hamiltonian  function $H$.

\medskip

Now, given two locally Hamiltonian $n$-vector fields 
it might seem  natural to  define
their bracket $[\ ,\ ]$ 
by means of the equality
\beq
\mbox{\nbrpq{n}{n} \inn\ \oo\, := \lieni{n}{1}(\xxi{n}{2}\inn\ \oo) , }
\eeq
which is just an extension of the definition of the Lie bracket of vector 
fields. From this definition 
it formally follows  that 
\beq
\mbox{$d^{V}$(\nbrpq{n}{n} \inn\ \oo) = 0,}
\eeq
so that \nbrpq{n}{n}\ is also locally Hamiltonian.  
However, the
bracket above does not map $n$-vectors  to $n$-vectors; 
instead it mixes
multivectors of different degrees. 
Moreover,  the counting of
degrees in eq. (3.7) gives $deg$(\nbrpq{n}{n}) = $2n-1$, 
so that the bracket identically vanishes  
unless $n=1$.  
This simple consideration 
indicates that multivector fields  and, consequently, 
forms of various  degrees should come into play. 
We are thus led to the following  construction.

\medskip

Given the polysymplectic $(n+1)$-form $\Om$, 
we  
define 
the set of {\em locally Hamiltonian} ($LH$) multivector fields 
\xx{p}, $1\leq p \leq n$,   
which satisfy the condition   
\beq 
\mbox{$\pounds$}_{\small \xx{p}}   
\Om  = 0.  
\eeq
The  $p$-vector field is then called {\em Hamiltonian} if there
exists a {\ horizontal} $(n-p)$-form $\ff{n-p}$
such that 
\beq
\mbox{\xxi{p}{F}\inn\ \oo \, = $ d^V$\ff{n-p},}
\eeq
The multivector field $\xx{p}$$_{F}$ is said to be 
the {\em Hamiltonian multivector field} associated with the
form $\ff{n-p}$.   
The horizontal forms $\ff{q}$ to which a Hamiltonian 
multivector field  can be associated 
are to be refered to as  {\em Hamiltonian forms}. 
Hamiltonian forms of various degrees 
extend to field theory, 
within the present approach, 
Hamiltonian functions, or dynamical variables,  in mechanics. 
The inclusion of forms of various degrees is motivated also by
the fact that the dynamical variables of interest in  
a field theory in 
$n$ dimensions can be represented in terms of horizontal 
forms of various degrees  $p \leq n$ 
($n$-forms are to be included in Sect. 5).   
It should be noted, however,  that, 
in contrast with the symplectic formulation of mechanics, 
the map in eq. (3.10) implies a rather strong restriction on
the allowable  dependence of the components of Hamiltonian forms 
on the polymomenta 
(see e.g. eq.  (4.5) below for the case of $(n-1)$-forms).
This limits the class of admissible Hamiltonian forms.  

Note that the map in (3.10) 
is invariant. In fact, under a coordinate transformation 
the vertical multivector $\xx{n-p}{}_F$ may gain higher vertical 
contributions (let us denote them $\xx{n-p}{}^{VV}$), 
the polysymplectic form may gain an $n$-horizontal part $W \we \vol$ 
in which $W$   is essentially 
a vertical covector,   
and the vertical exterior differential of $\ff{p}$ may gain a 
pure $(p+1)$-horizontal addition 
$\raisebox{1pt}{$\stackrel{p+1}{f}$}$. 
Thus, in another coordinate system 
(3.10) takes the form 
$(\xx{n-p} + \xx{n-p}{}^{VV} )  \inn\ (\Omega + W\we \vol ) 
= \dv \ff{p} + \raisebox{1pt}{$\stackrel{p+1}{f}$} $  
whence it follows: 
$\xx{n-p} \inn\ \Omega = \dv \ff{p} $ and  
 $\xx{n-p}\inn (W\we \vol) +  \xx{n-p}{}^{VV}\inn\ \Omega = 
\raisebox{1pt}{$\stackrel{p+1}{f}$} $. 
The former equation coincides with (3.10), 
while the latter, being pure horizontal, 
is factored off,  
for (3.10) is understood as an equality of equivalence classes 
modulo the horizontal contributions of highest degree. 

\medskip

The bracket of two locally Hamiltonian fields may 
now be defined by
\beq 
\mbox{\nbrpq{p}{q}\inn\ \oo\ } := 
\mbox{$\pounds$}_{\mbox{\small $\xx{p}{}_1$}}(\xx{q}{}_2 
\inn\ \Omega).  
\eeq
It is easy to show that the bracket defined above  
maps the pair of {\em LH} fields 
to a Hamiltonian field and that  
\begin{quote}
(i) it generalizes the Lie bracket of vector fields,\\
(ii) its multivector degree is 
\end{quote}
\beq
\mbox{$deg($\nbrpq{p}{q}$) = p+q-1$, }
\eeq
\begin{quote}
(iii) it is graded antisymmetric
\end{quote}
\beq
\mbox{\nbrpq{p}{q} = $-(-1)^{(p-1)(q-1)}$ \nbr{$\xxi{q}{2}$}{$\xxi{p}{1}$}}, 
\eeq
and 
\begin{quote}
(iv) it fulfils the graded Jacobi identity 
\end{quote}
\begin{eqnarray}
\mbox{$(-1)^{g_1 g_3}$\nbr{\xx{p}}{\nbr{\xx{q}}{\xx{r}}}} &+&   
                                           \mbox{\hspace*{15em}} \nonumber \\
\mbox{$(-1)^{g_1 g_2}$\nbr{\xx{q}}{\nbr{\xx{r}}{\xx{p}}}} &+&\mbox{$
 (-1)^{g_2 g_3}$ \nbr{\xx{r}}{\nbr{\xx{p}}{\xx{q}}} $=0,$}  
\end{eqnarray}
where $g_1=p-1, \; g_2=q-1$ and $g_3=r-1$.

All the properties above are known for the 
Schouten--Nijenhuis (SN) \cite{SN} bracket of multivector fields.
This is not of surprise, as the relation between the notion 
of Lie derivative with respect to a multivector field 
and the SN bracket is already evident from  \cite{tulcz} 
(see also a related discussion in \cite{Dolan}). 
However, in our case there is a little 
difference  due to the fact that  
the bracket is defined on vertical multivectors, 
so that it contains only the derivatives 
with respect 
to the vertical variables, 
while the multivectors themselves 
have both vertical and horizontal indices.
Despite this  difference, the bracket of 
locally  Hamiltonian multivector fields defined in (3.11) 
will be referred to here as the (vertical) SN bracket.
It is clear from the above equations that the set of $LH$ multivector
fields equipped with the SN  bracket  
is a {$\Bbb Z$}-graded Lie algebra.     

Now, taking \xxi{p}{1}\ and \xxi{q}{2} to be Hamiltonian multivector 
fields, we obtain: 
\begin{eqnarray}
\mbox{\nbrpq{p}{q}\inn\ $\Omega$} & =\, &  
                 \mbox{\lieni{p}{1}$d^{V}$\ff{s}$_{2}$} \nonumber \\ &=\, &  
                 \mbox{$ (-1)^{p+1}\, d^{V}($\xxi{p}{1}\inn\
$d^{V}$\ff{s}$_{2})$} \nn \\ 
& =:  & -
d^{V}\pbr{\ff{r}_{1}}{\ff{s}_{_2}}, 
\end{eqnarray}
where $r=n-p$ and $s=n-q$. 
The first equality in (3.15) follows from (3.10) and (3.11). 
The second one essentially proves that 
the SN bracket of two Hamiltonian 
multivector fields is also a Hamiltonian multivector field.  
The third equality 
identifies the form which is associated 
with the SN bracket of two Hamiltonian multivector fields 
with the  bracket of forms which are associated with these 
multivector fields. 

We shall call the bracket operation on forms 
introduced in (3.15) 
the {\em graded Poisson bracket}. 
In the following subsection it will be shown   
that it fulfills a certain graded generalization 
of the properties of the usual Poisson bracket. 
It is also evident from the definition 
that our graded Poisson bracket of forms  
is related to the Schouten-Nijenhuis bracket 
of  multivector Hamiltonian fields 
and the polysymplectic 
form in just  
the same way as the usual
Poisson bracket of functions is related 
to the Lie bracket of Hamiltonian vector fields 
and the symplectic form.

From eq. (3.15) the following 
useful formulae for the graded Poisson bracket of forms
can be deduced 
\beqa
\pbr{\ff{r}_1}{\ff{s}_2} 
&=& (-1)^{(n-r)} \mbox{\lieni{n-r}{1} \ff{s}$_{2}$} 
\nn \\
 &=& (-1)^{(n-r)}\xx{n-r}{}_{1} \inn\ d^{V} \ff{s}_2 
 \\
 &=& (-1)^{(n-r)}\xx{n-r}_{1} \inn\ \xx{n-s}_{2} \inn\ \Omega.  \nn 
\eeqa 
These relations generalize  the usual definitions of 
the Poisson bracket in mechanics, 
but they are merely a consequence, 
as in mechanics, of the more  fundamental definition, eq. (3.15), 
which is directly related 
to the basic graded canonical symmetry principle, eq. (3.9). 
Note, that in spite of the fact
that the definition in eq. (3.15) determines only 
the vertical exterior differential  of the Poisson bracket,  
there is no arbitrariness 
"modulo an exact form" 
in the definition of the Poisson bracket itself 
(as opposed to the previous 
definitions of the Poisson bracket of $(n-1)$-forms, 
cf. \cite{Gold+Stern,Herm lie,Kij ea}) 
because the latter maps
horizontal forms to horizontal forms, 
while the  $d^V$-exact addition 
would be  necessarily vertical (if understtod as a coset). 
Note also that 
using a consideration similar to that which demonstrated 
the invariance of the map (3.10) 
the invariance  of the graded Poisson bracket given by (3.16) 
can be established.

\subsection{\normalsize\bf Algebraic properties of  graded Poisson bracket} 

Let us consider the algebraic properties of 
the graded Poisson
bracket of forms.
The degree counting in eq. (3.15) gives
\beq
deg\pbr{\ff{p}_1}{\ff{q}_2} = p+q-n+1 , 
\eeq 
so that  the bracket exists if $p+q\geq n-1$.
From the graded anticommutativity of the SN bracket 
the similar  property can be deduced  for the graded Poisson bracket:
\beq
\pbr{\ff{p}_1}{\ff{q}_2} = -(-1)^{g_1 g_2}
\pbr{\ff{q}_2}{\ff{p}_1}, 
\eeq
where  $g_1 := n-p-1$ and $g_2 := n-q-1$ are 
degrees of corresponding forms with respect to the bracket operation. 
Furthermore, by a straightforward calculation 
the graded Jacobi identity can be proven:
\begin{eqnarray}
\mbox{$(-1)^{g_1 g_3} \pbr{\ff{p}}{\pbr{\ff{q}}{\ff{r}}}$}  &+& \nn \\
\mbox{$(-1)^{g_1 g_2} \pbr{\ff{q}}{\pbr{\ff{r}}{\ff{p}}}$} &+& 
\mbox{$(-1)^{g_2 g_3} \pbr{\ff{r}}{\pbr{\ff{p}}{\ff{q}}}= 0,$}  
\end{eqnarray}
where $g_3 := n-r-1$.

\medskip

Thus, 
the space
of Hamiltonian forms equipped with 
the graded Poisson bracket operation  
defined in (3.15), 
is  a {$\Bbb Z$}-graded Lie algebra. 
Now, the question naturally arises as 
to whether this bracket 
gives rise 
also to some appropriate analogue of the Poisson algebra structure, 
as is the case in mechanics. 

In order to answer this question,
the analogue of 
the Leibniz rule has to be considered. 
It should be noted from the very beginning that 
the space of Hamiltonian forms 
on which the bracket $\pbr{.}{.}$ has so far been defined 
is not stable with respect 
to the exterior product. 
For example, if an $(n-1)$-form $\ff{n-1}$ is Hamiltonian, then it 
is easy to show 
that its product with 
any function 
which depends on the polymomenta is not a Hamiltonian 
$(n-1)$-form because  Hamiltonian $(n-1)$-forms must have a 
very specific (linear) dependence on the polymomenta 
(cf.  eq. (4.5)). 
Therefore, there is little sense in the Leibniz rule with 
respect to the exterior product within 
the space of Hamiltonian forms. 
However,  the 
dynamical variables 
in field theory which 
cannot be represented as Hamiltonian forms can be easily constructed. 
For example, it is  quite 
natural to associate 
the $(n-1)$-form $P_j := T^i_j\, \der_i\inn \vol$
to the canonical energy-momentum tensor 
$T^i_j := p_a^i \der_j y^a - \delta^i_j L$. 
It 
is easy to see   
that expressing it in terms of the DW variables so that  
$T^i_j = T^i_j (y^a, p^i_a, x^i)$ one is led to the $(n-1)$-form $P_j$ which 
is not Hamiltonian for most  field theories that are of interest. One may 
expect, therefore, that the algebraic structure on Hamiltonian 
forms is embedded in 
some more general structure which does also involve non-Hamiltonian forms. 
Investigation of the validity of the Leibniz rule with respect to the exterior
product may be considered as an attempt to gain 
insight into this more 
general structure (see also the related discussion at the end of 
Sect. 5). An alternative tactic which is not considered  here 
might be a construction of the multiplication
law of horizontal forms 
with respect to which the space of Hamiltonian forms 
is stable.\footnote{See the Note at the end of Sect. 5.}

Let us consider first the left Leibniz rule. 
The calculation in the Appendix  
yields the following result: 
\beqa
&&\pbr{\ff{p}}{\ff{q} \wedge \ff{r}} = 
\pbr{\ff{p}}{\ff{q}} \wedge 
\ff{r} + (-1)^{q(n-p-1)} \ff{q} \wedge \pbr{\ff{p}}{\ff{r}} \nn \\ 
&& \nn \\
&-&(n-p)(-1)^{n-p} \sum_{s=1}^{n-p-1} 
\xx{n-p}{}^v{}\uind{i}{s}{}^{i_{s+1} ... i_{n-p-1}}
[(-1)^{q(n-p-s-1)}
\der^{\otimes}\lind{i}{s} 
\inn \ff{q}) 
\eeqa
\[ \we \, 
\der^{\otimes}_{i_{s+1} ... i_{n-p-1}}
\inn\der_v\inn 
\dv \ff{r}  
+ (-1)^{s(n-p-q-s-1)} ( \der^{\otimes}_{i_{s+1} ... i_{n-p-1}} 
\inn\der_v\inn 
 \dv \ff{q}) \we 
\der^{\otimes}\lind{i}{s} 
\inn \ff{r}) , \]      
where $\xx{n-p}$ is the Hamiltonian multivector field associated with 
$\ff{p}$ and
$\der^{\otimes}\lind{i}{p} 
:= \der_{i_1}\otimes ... \otimes \der_{i_p}$ .

The first two terms in (3.20) 
are typical for the graded Leibniz 
rule. 
However, the supplementary terms are also present, 
appearing due to the fact that the 
multivector field of degree $(n-p)$ 
is not a graded derivation on the exterior algebra, 
but rather a kind of  graded 
differential operator of order $(n-p)$ 
(see App. 1 for more details). 
If there were  no additional higher order terms  
in the last two lines  of eq. (3.20),
 a {$\Bbb Z$}-graded
Poisson algebra with different gradings with respect to the graded 
commutative exterior product and the graded anticommutative 
bracket operation would be arrived at. This structure is   
known as a Gerstenhaber algebra \cite{Gerstenhaber}.
However, the expression obtained here 
is essentially a higher-order analogue of the graded Leibniz rule 
(see  App. 1), 
so that the algebraic structure 
which emerges here may be viewed 
as a {\em higher-order} generalization of  Gerstenhaber algebra. 
This notion means that the bracket operation acts as a higher-order 
graded differential operator on a graded commutative algebra 
instead of being a graded derivation.    

The above formulation  of the higher-order Leibniz rule  is 
not quite satisfactory 
as it is not given entirely in terms of 
the bracket with $\ff{p}$ and 
refers explicitly to the 
components of the multivector field  associated with the form $\ff{p}$. 
More appropriate  formulation may be given in terms of the 
$r$-linear maps $\Phi^r_D$ introduced by 
Koszul \cite{koszul}, 
which are associated with 
a graded higher-order 
differential operator $D$ on a graded commutative algebra. 

\newcommand{\Lm}{\bigwedge_0^*}

On the space of horizontal forms $\Lm$ the map 
$\Phi{}^r_D: \bigotimes{}^r \Lm \rightarrow \Lm$ is 
defined by 
\beq
\Phi{}^r_D(F_1, ..., F_r) := 
m \co (D\otimes {\mbox{\bf 1}}) 
\lambda{}^r (F_1\otimes ... \otimes F_r),
\eeq
for all $F_1, ..., F_r$ in $\Lm$.
Here $m$ denotes the multiplication map: 
\[
m(F_1\otimes F_2):=F_1 \we F_2 ,
\]
$\lambda{}^r$ is a linear map 
$\bigotimes{}^r \Lm \rightarrow \Lm \otimes \Lm$ such that 
\[
\lambda{}^r (F_1\otimes ... \otimes F_r) :=
\lambda(F_1) \we ... \we \lambda(F_r)  ,
\]
and $\lambda: \Lm \rightarrow \Lm \otimes \Lm $ 
is the map given by 
\[
\lambda(F) := F\otimes {\mbox{\bf 1}} - {\mbox{\bf 1}} \otimes F .
\]
The graded differential operator $D$ is said to be of $r$-th 
order iff $\Phi{}^{r+1}_D = 0$ identically. 
This definition may be checked to be  consistent with the idea 
of $r$-th order partial derivative on the multiplicative 
algebra of functions. 

Now, the higher-order Leibniz rule
fulfilled by the (right) bracket with a $p$-form
may be written compactly as follows:
\beq
{\mbox{\Large $\Phi$}}^{n-p+1}_{\mbox{\small $\pbr{\ff{p}}{\,.\,} $} }
(F_1, ... , F_{n-p+1})=0.
\eeq
For $p=(n-1)$ this reproduces the usual Leibniz rule, because the 
vector field associated with a form of degree $(n-1)$ is a derivation on
the exterior algebra. The simplest non-trivial example is obtained when
$p=(n-2)$. This leads to the second-order graded Leibniz rule 
\[
{\mbox{\Large $\Phi$}}^{3}_{\mbox{\small $\pbr{\ff{n-2}}{\,.\,} $} }
(\ff{q},\ff{r},\ff{s})=0 
\]
or, in the explicit form, 
\beqa
\pbr{\ff{n-2}}{\ff{q}\we\ff{r}\we\ff{s}}
&=&\pbr{\ff{n-2}}{\ff{q}\we\ff{r}}\we\ff{s} 
+(-1)^{q(r+s)}\pbr{\ff{n-2}}{\ff{r}\we\ff{s}}\we\ff{q} 
\nn \\
&+& (-1)^{s(q+r)}\pbr{\ff{n-2}}{\ff{s}\we\ff{q}}\we\ff{r}
- \pbr{\ff{n-2}}{\ff{q}}\we\ff{r}\we\ff{s}
\\
&-&(-1)^{q(r+s)}\pbr{\ff{n-2}}{\ff{r}}\we\ff{s}\we\ff{q}
- (-1)^{s(q+r)}\pbr{\ff{n-2}}{\ff{s}}\we\ff{q}\we\ff{r} ,
\nn
\eeqa
where the  expression of $\Phi{}^3_D$ as found in \cite{koszul} 
is used. One can better understand the expression above by comparing it 
with the ``Leibniz rule for the second derivative'' 
\[
(abc)''=(ab)''c+(ac)''b+(bc)''a-a''bc-ab''c-abc''.
\]

The structure defined by 
eqs. (3.18), (3.19), (3.22) and the graded 
commutativity of the exterior product  generalizes both the 
Poisson algebra structure on functions and the Gerstenhaber algebra structure 
on graded commutative algebra. We suggest   
to call this structure a 
{\em higher-order Gerstenhaber algebra}.  

\medskip

Now, let us consider the right Leibniz rule. In this case we are 
interested in the expression 
\beq
\pbr{\ff{q}\we\ff{r}}{\ff{p}} 
= (-1)^{n-q-r}
\xt_{\mbox{\small $\ff{q}\we\ff{r}$ }} 
\inn \dv \ff{p} , 
\eeq
where the symbol $\xt_{\mbox{\small $\ff{q}\we\ff{r}$ }}  $ 
denotes the object which is associated with the form 
$\ff{q}\we\ff{r}$. Because the exterior product of two 
Hamiltonian forms is not necessarily Hamiltonian this 
object is not generally a multivector. Still, we can attribute  
a certain meaning to it as follows. By definition,
\[
\xt_{\mbox{\small $\ff{q}\we\ff{r}$ }} 
\inn\ 
\Omega = \dv(\ff{q}\we \ff{r}), 
\]
where the right hand side may be written as follows  
\beqa
\dv\,(\ff{q}\we \ff{r}) &=&(-1)^{r(q+1)}\ff{r}\we 
\dv\ff{q}+ (-1)^q \ff{q}\we  \dv\ff{r}  
\nn \\
&=& (-1)^{r(q+1)}\ff{r}\we X_{\ff{q}}\inn \Omega 
+ (-1)^q \ff{q}\we X_{\ff{r}}\inn 
\Omega . 
\nn
\eeqa 
Hence, we can formally take 
\beq
\xt_{\mbox{\small $\ff{q}\we\ff{r}$ }}  := 
(-1)^{r(q+1)}\ff{r}\co X_{\ff{q}} 
+ (-1)^q \ff{q}\co X_{\ff{r}} ,  
\eeq
where the right hand side is understood as a 
composition of two operations, the inner product with a multivector, 
and the exterior product with a form. 

With this definition the Poisson bracket of interest takes 
the form 
\beq
\pbr{\ff{q}\we\ff{r}}{\ff{p}} =
\ff{q}\we\pbr{\ff{r}}{\ff{p}}
+(-1)^{rq}\ff{r}\we\pbr{\ff{q}}{\ff{p}} ,
\eeq
so that the right graded Leibniz rule 
appears to be fulfilled. 
The algebraic structure defined by 
the graded Lie algebra properties of the bracket, 
eqs. (3.18) and (3.19), 
graded commutativity of the exterior product, 
and the right graded Leibniz rule (3.26) 
is known as a {\em right} Gerstenhaber algebra 
(see e.g. \cite{flato} for the explicit definition). 

Thus, evidence of two possible ways 
in which a generalization of the  
(graded) Poisson structure is introduced 
on the space of horizontal forms 
by the bracket operation defined in (3.15)
is found. 
The first  leads to the new notion of  higher-order 
Gerstenhaber algebra, while  the second provides us with the 
structure of right Gerstenhaber algebra. Note however, that 
both structures 
are based on a certain extrapolation beyond the space of Hamiltonian 
forms, and their adequate substantiation requires a proper definition of the 
bracket operation on arbitrary horizontal forms. 
The formal construction of the object in (3.25) 
associated with  the exterior product of two forms 
suggests that non-Hamiltonian forms can be associated with 
multivector valued forms. 
Note also that the puzzling  
difference between left and right Leibniz rules, 
as well as the subsequent consideration
in Sect. 5 
of the bracket with $n$-forms  
which appears  
to be not  graded antisymmetric in general,  
poses the question 
whether the graded antisymmetry 
of the  Poisson bracket of forms 
can be preserved for non-Hamiltonian forms.

\subsection{\normalsize\bf Pre-Hamiltonian  fields}

In the present formulation  
there exists  a nontrivial set of Hamiltonian 
multivector  fields \xxi{p}{0} which annihilate the polysymplectic form:
\[\xxi{p}{0} \inn\ \Omega := 0,       \]
$p=1,...,n.$ 
We shall call them {\em pre-Hamiltonian}  fields.\footnote{In our 
earlier papers the term ``primitive Hamiltonian fields'' have been 
used instead.}  
As a consequence of their existence 
the map in eq. (3.10) specifies a Hamiltonian multivector field 
associated with a given form 
only 
up to an addition of a pre-Hamiltonian field. 
This means that it actually 
maps Hamiltonian $(n-p)$-forms
to the {\em equivalence classes} $[\xx{p}]$ 
of Hamiltonian multivector fields of degree $p$  
modulo an addition of  pre-Hamiltonian 
$p$-vector fields: 
$[\xx{p}]=[\xx{p}+\xx{p}$$_{0}]$.
It is easy to show that pre-Hamiltonian fields  
form an ideal   $\cal X$$_{0}$ in 
the graded Lie algebra $\cal X$ of Hamiltonian multivector fields 
with respect to the SN bracket, 
and that the graded Poisson bracket in (3.16) does not depend on the 
choice of the representative of the equivalence class of multivectors 
associated with $\ff{r}$ or $\ff{s}$.  
Therefore the graded quotient algebra 
$\cal X$/$\cal X$${}_{0}$ 
is essentially  
a field theoretical analogue  
of 
a 
Lie algebra of Hamiltonian 
vector fields in mechanics.  
Note also that 
this is on the quotient space of 
Hamiltonian multivector fields modulo pre-Hamiltonian fields
the polysymplectic form may be considered as non-degenerate.

\section{\normalsize\bf 
The equations of  motion  of 
Hamiltonian   
 (n--1)-forms 
 }

In this section,  the operation on 
Hamiltonian forms
generated by the graded  Poisson bracket with the DW Hamiltonian
function is considered, 
and the equations of motion of 
Hamiltonian $(n-1)$-forms are 
written 
with the help of this bracket. 
Next, the field theoretic analogues  
within the polymomentum formulation   
of the integrals of the motion and 
the canonically conjugate variables are  discussed. 

{}From the degree counting in eq. (3.16) 
it is evident that only the forms of degree $(n-1)$
have nonvanishing brackets with $H$ 
and that the resulting brackets 
are  forms of degree $0$. Let us
calculate the bracket of 
a Hamiltonian $(n-1)$-form
\[F:= F^{i}\der_{i}\inn\ \widetilde{vol}\]
with the DW Hamiltonian function $H$:
\beq
\pbr{F}{H} = - \xx{}_{F}\inn\ d^{V}H.  
\eeq
The components of the vector field associated with $F$, 
\[ \xxi{}{F} := X^a\der_a + X^i_a\der^a_i , \] 
are given by 
the equation
\[ \xxi{}{F} \inn\ \Omega= d^{V} F \]  
or in components 
\beq
 \mbox{$(-X^a dp^i_a + X^i_a dy^a)\wedge \dvol{i} 
= (\der_a F^i dy^a + \der ^a_j F^i dp^j_a)\wedge \dvol{i}$ } . 
\eeq
{}From (4.2) we obtain 
\beq
X^{a} \delta^{i}_{j}=-\der^{a}_{j}F^{i},
\eeq
\beq
X^{i}_{a}=\der_{a}F^{i}.
\eeq
We see that   
no arbitrary $(n-1)$-forms   
can be Hamiltonian, that is to say, 
no arbitrary $(n-1)$ forms can 
ensure the consistency of both sides of (4.2)
and  be associated with a certain  Hamiltonian vector field, but only those
which satisfy the condition (4.3) which restricts the admissible 
dependence of the
components of Hamiltonian $(n-1)$-form   on polymomenta $p^i_a$.  
Namely, from the integrability condition of 
(4.3) we can deduce  
that the most general admissible
Hamiltonian $(n-1)$--form has the components
\beq
F^i = - p_a^i X^a(y,x) + g^i(y,x), 
\eeq 
where $X^a$ and $g^i$ are arbitrary functions of 
field and space-time variables. 
For such
$(n-1)$-forms, 
from  (4.1) and  (4.3) - (4.5)  
it follows
\beq
\pbr{H}{F} = \der_a F^i \der^a_i H + X^a \der_a H.
\eeq 
Now, let us introduce the {\em total\/} 
(i.e. evaluated  on sections $z^v=z^v(x)$)   
exterior differential
$\bd$ of the form {\em F} 
\[\mbox{{\bd}}F:= (\der_a F^j \der_i y^a + \der^a_k F^j \der_i p^k_a 
+ \der_i F^j ) dx^i \wedge \dvol{j} .\] 
Taking into account the condition (4.3) yields 
\[\mbox{{\bd}}F= (\der_a F^i \der_i y^a - X^a \der_i p^i_a 
+ \der_i F^i) \vol . \]
Using 
the DW Hamiltonian 
field equations, eqs. (2.3), in the equation above, 
and comparing the result with (4.6), 
for an arbitrary Hamiltonian $(n-1)$-form $F$ 
we obtain  
\beq
\mbox{{\bd}$F$} =  \pbr{H}{F} \widetilde{vol} + d^{hor} F, 
\eeq
where the last term 
$d^{hor} F  = (\der_{i}F^{i})\,\widetilde{vol} \,$ 
appears for forms which have 
an 
explicit dependence on the  space-time variables.  
Taking the inverse Hodge dual of  
(4.7)\footnote{Recall that
$\star \vol = \sigma$, where $\sigma=+1$ for Euclidean and $\sigma=-1$
for Minkowski signature of the metric; $\star^{-1}\, \star := 1$,
therefore on $n$-forms $\star^{-1} =
\sigma \star$ or, in general, $\star^{-1} = \sigma (-1)^{p(n-p)} \star$ 
on $p$-forms.}: 
\beq
\mbox{$\star$$^{-1}${\bd}$F =$} \pbr{H}{F} + \der_{i}F^{i}, 
\eeq
we 
conclude that the Poisson bracket of a Hamiltonian $(n-1)$-form 
with the DW Hamiltonian function is 
related to the inverse Hodge dual of the total exterior differential 
of the former. Eq. (4.8) is the equation of motion of 
a  Hamiltonian form of degree $(n-1)$  
in Poisson bracket formulation. 
It obviously extends the familiar Poisson bracket formulation of the 
equations of motion in mechanics to the DW formulation of field theory. 
Note, that the dual of the total exterior derivative
$\star^{-1}$\mbox{\bd} of $(n-1)$-forms 
can be related to the generalized Lie derivative 
with respect to the $n$-vector field 
annihilating the canonical $(n+1)$-form $\Omega_{DW}$   
(see eqs. (2.6)--(2.8)) making the analogy with 
the equations of motion in mechanics even more transparent. 
In the subsequent section we  discuss how  the 
bracket representation of the equations of motion 
can be generalized to
forms of degrees $p\leq(n-1)$.

\medskip 

\subsection{ 
DW canonical equations in Poisson bracket formulation 
and canonically conjugate variables }

Eq. (4.8) contains, as a special case, the entire set of 
DW Hamiltonian field equations, eqs. (2.3). 
In fact,  on account of $d^{hor} y^a
= 0$ and $d^{hor} p^i_a = 0$, 
by substituting $p_a := p^i_a \dvol{i}$
and then $y^a_i := y^a \dvol{i}$ 
for the $(n-1)$-form $F$ in (4.8), 
using (4.6) we obtain
\beqa
\mbox{$\star^{-1}${\bd }} p_a &=& \pbr{H}{p_a}\; = \;- \der_a H , \nn \\
&& \\
\mbox{$\star^{-1}${\bd }} y^a_i &=& \pbr{H}{y^a_i}\; = \;\der^a_i H. \nn
\eeqa
which is the Poisson bracket formulation of 
DW Hamiltonian field equations, eqs. (2.3).  

The representaion above of the canonical field equations 
in terms of the graded Poisson bracket  
sheds also light on the question 
as to which variables may be considered as canonically conjugate 
within the formalism under discussion. 
As we know from mechanics,
the canonically conjugate variables 
have  "simple" mutual Poisson
brackets (leading to the Heisenberg algebra structure) and 
their product has the dimension of action.  

It is easy to see that in our
approach the pair of variables
\beq
y^a \quad {\rm and } \quad p_a := p^i_a \dvol{i}),
\eeq
one of which is 
a $0$-form and another 
is an $(n-1)$-form, 
may be considered as a pair of canonically conjugate variables,  
for     
the Poisson brackets of these variables
\beq
\pbr{y^a}{p_b} = - \delta^a_b, \quad 
\pbr{y^a}{y^b}=0, \quad 
\pbr{p_a}{p_b}=0
\eeq  
agree with those of coordinates and canonically
conjugate momenta in mechanics. 

Indeed, from  
(3.16) we deduce  
\[\pbr{y^a}{p_b}=-\pbr{p_b}{y^a}=\mbox{\xx{1}$_{(p_b)}$} \inn\, dy^a = 
-\delta^a_b, \] 
where  in the last equality 
we used the fact that the  
vector field \xx{1}$_{(p_b)}$ associated with the $(n-1)$-form $p_b$ is
given by
\[\mbox{\xx{1}$_{(p_b)}$} \inn\, \Omega = dp_b , \]
whence it follows that 
\[\mbox{\xx{1}$_{(p_b)}$}=-\der_b . \]

It should be noted, however, that
the choice above of the pair of canonically conjugate variables  
is not unique.  For example, one  could also choose the pair 
$(y^a \dvol{i},p^j_b)$   
or $(y^a \dvol{i},p_b) $ 
since in this case  
the (non-vanishing) Poisson brackets are 
also similar to the canonical bracket in mechanics and 
reduce to the latter when $n=1$:          
\beq
\pbr{y^a \dvol{i}}{p^j_b} = - \delta^j_i\, \delta^a_b , 
\quad   
\pbr{y^a \dvol{i}}{p_b} = - \delta^a_b \, \dvol{i}. 
\eeq
Such a freedom 
in specifying the pairs of canonically conjugate variables 
is due to the "canonical graded symmetry", eq. (3.9), 
mixing
forms of different degrees. 
It may be especially useful in field
theories where the field variables themselves are forms. 
For example, 
the 1-form potential $A_{i}dx^{i}$ in electrodynamics or 
the  2-form
potential $B_{ij}dx^i\we dx^j$ 
in the Kalb-Ramond 
field theory.
In general, 
an 
$(n-p-1)$-form conjugate momentum may be associated
to a $p$-form field
variable so  that their mutual Poisson bracket 
is a constant equal to one when $n=1$ 
(cf. Sect. 6.2 where the example of electrodynamics is 
considered).   


\medskip 
        
\subsection{\normalsize\bf The conserved currents}

The Poisson bracket formulation of the equations of motion 
suggests a natural generalization to field theory 
of the classical notion of an integral of the motion known 
in analytical mechanics.  Let $\cal J$ be a Hamiltonian $(n-1)$-form which does
not depend explicitly on 
space-time coordinates and whose Poisson bracket  
with the DW Hamiltonian function vanishes. Then, from eq. (4.8)
the {\em conservation law} follows: 
\[\mbox{{\bd} $\cal J$$=0$.}\]
Therefore, the field theoretical analogues of integrals of the motion 
within the present formulation 
are  $(n-1)$-forms corresponding to
the conserved currents. 
Similar to the conserved quantities in mechanics, they are
characterized by the condition
\beq 
\pbr{{\cal J}}{H} = 0.
\eeq
Taking $\cal J$$_1$ and $\cal J$$_2$ to be  the 
$(n-1)$-forms satisfying
eq. (4.10) and using the Jacobi identity for the bracket 
we obtain 
\[ \pbr{H}{\pbr{{\cal J}{}_1}{{\cal J}{}_2}}=0.\]
Therefore, the Poisson bracket of two conserved currents
is again a conserved current.
The latter statement
extends to field theory 
the {\em Poisson theorem} \cite{Arnold}, 
to the effect that
the Poisson bracket of two integrals of the motion 
is again an integral of the motion. 
One can also conclude that the set of
conserved $(n-1)$-form currents having a vanishing Poisson  bracket with
the DW Hamiltonian function is closed with respect to the Poisson
bracket and therefore forms 
a Lie algebra which is a 
subalgebra of graded Lie algebra of  Hamiltonian forms.

Furthermore, eq. (4.10) means that the Lie derivative of $H$ 
with respect to  the 
vertical  vector field $X_{\cal J}$ associated with $\cal J$ 
vanishes, i.e. $H$
is invariant with respect to the symmetry generated 
by $X_{\cal J}$, and 
$\cal J$ is the 
conserved current corresponding to this symmetry of the DW Hamiltonian.
This is obviously a 
field theoretical extension (within the present framework) 
 of the {\em Hamiltonian Noether theorem} 
(cf. for example Ref. \cite[a]{Arnold} $\S$40
or Ref. \cite[b]{Arnold} $\S$15.1). 
Note that this extension is proved here
for the symmetries  generated by  {\em vertical} vector fields, 
that is  only for  internal symmetries.    
Another considerable limitation of the present discussion is 
that $(n-1)$-forms ${\cal J}$ are supposed to be 
Hamiltonian, 
that is restricted to have the form (4.5).


\section{\normalsize\bf The equations of motion of 
forms of arbitrary degree} 

   In Sect. 3 it was argued that  the proper field theoretical
generalization of  Hamiltonian functions 
or dynamical variables  are 
horizontal forms of various degrees from $0$ to $(n-1)$, 
on the subspace of which given by the Hamiltonian forms 
a graded analogue of the Poisson bracket operation 
can be defined. 
However, in Sect. 4 only the equations of
motion of Hamiltonian forms of degree $(n-1)$ were 
formulated in terms of the graded Poisson bracket.  
It seems natural to ask whether this circumstance is
due to some privileged 
place of  $(n-1)$-forms in the present formalism
(which might indeed be the case since 
these forms 
yield classical observables of the field 
after integrating over the 
space-like hypersurface)   
or a possibility exists to
represent  the equations of motion of 
forms of any  degree in the bracket formulation. 
We show in this section that the second alternative 
may indeed be  realized.  
This requires, however,  a slight generalization of the construction
presented in Sect. 3.  
In particular, an extension of the notions of a Hamiltonian
form and the associated Hamiltonian multivector field is required.  

The problem faced 
when trying to extend the Poisson bracket formulation of 
the equations of motion from 
$(n-1)$-forms to 
forms of arbitrary degree is 
that the bracket $\pbr{\ff{p}}{H}$ 
vanishes  identically when $p<(n-1)$. 
A  way out is suggested by the observation
that for all $p$ the bracket with the $n$-form $H\vol$ 
would not vanish, if properly defined, 
as the formal degree counting based on eq. (3.17) 
indicates that its degree is to be $(p+1)$. 
In order to define such a bracket 
our hierarchy of equations relating  Hamiltonian forms to 
Hamiltonian multivector fields, eq. (3.10), 
has to  be extended to 
horizontal forms of degree $n$ 
so that 
a certain object  generalizing  Hamiltonian multivector field 
could be associated with the form $H\vol$. 
It is easily 
seen that the formal multivector degree of such an object 
has to be equal to zero.

This is possible indeed, if the object
$\xt{}^V$ which is associates with the horizontal $n$-form
$\ff{n}$ by means of the map
\beq
\mbox{$\tilde{X}$$^{V}_{F}$ \inn\ }\Omega = d^{V} \ff{n}  
\eeq
is 
a vertical-vector valued horizontal one-form 
\beq 
\tilde{X}^{V} := X^{v}_{\cdot\, k}\, dx^{k}  \otimes  
\der_{v}      
\eeq
which acts on the form to the right 
through the Fr\"{o}licher-Nijenhuis (FN) inner
product. Recall that the FN inner product 
of a vector valued form with a form is defined as follows
(see e.g. \cite{FN,Nono,kolar}) 
\beq
\tilde{X}{}^V \inn\ \Omega\ := X^{v}_{\cdot\, k} dx^{k} \wedge\, (\der_{v}
\inn\ \Omega).  
\eeq 
Here we continue to use the usual symbol 
of the inner product of vectors and forms
and imply that the tilde over the argument 
to the left indicates that
it is a vector valued form, so that the sign \inn\ 
respectively 
denotes the  FN
inner product.   

By extending formulae (3.16) 
to the case when one of the arguments in the Poisson bracket is 
an $n$-form   the 
bracket of the  $p$-form \ff{p}  with  
the $n$-form \ff{n} can be defined as follows
\beq
\pbr{\ff{n}_1}{\ff{p}_2}' 
: = \tilde{X}^{V}_{F_1} \inn\ d^V \ff{p}_2. 
\eeq 
This expression may be substantiated by considerations similar to
those which led from eq. (3.9) to eq. (3.16).   
To this end, the 
hierarchy of symmetries in eq. (3.9) 
has to be supplemented with the additional equation
\beq
\mbox{$\pounds$$_{\tilde{X}^{V}}$$\Omega = 0$ } 
\eeq 
which formally corresponds to $p=0$. 
The generalized Lie derivative 
with respect to the 
vertical-vector valued  form $\xt{}^{V}$  
is naturally given by (cf. (3.4))
\beq
\mbox{$\pounds$$_{\tilde{X}^{V}}$$ \Phi  := \tilde{X}^{V} \inn\ d^{V} \Phi -
d^{V}(\tilde{X}^{V} \inn\ \Phi)$  }  
\eeq
for an arbitrary form $\Phi$. Obviously, 
$\pounds$$_{\tilde{X}^{V}}$ maps  $p$-forms to $(p+1)$-forms.

It should be noted  that the bracket operation 
with an $n$-form as defined in (5.4) 
is not graded antisymmetric in general.  
That is why it is marked with a prime. 
Nevertheless,  
this is the definition 
which is shown below to be appropriate for 
representing the equations of motion of arbitrary horizontal 
forms in Poisson  bracket formulation. 

\medskip

Taking now $\ff{n} = H\vol$, let us calculate 
the components  of the associated 
vertical-vector valued form $\xt_{H\vol}$:
\beq
\mbox{$\tilde{X}$$^a_{\cdot k}$$ = \der^a_k H,$ \hspace*{1em} 
$\tilde{X}$$^i_{a k}$$\delta^k_i = -\der_a H. $}
\eeq 
It is evident that $\xt_{H\vol}$ is given 
up to 
an addition of an arbitrary (pre-Hamiltonian)  
vector valued form 
$\xt_{0}$ which satisfies
\[\tilde{X}{}_{0} \inn\ \Omega = 0 ,  \]
so that the only non-vanishing components of $\xt_{0}$  
obey  
$\xt_{0}{}^i_{ak} \delta^k_i =0$. 
Therefore, the symbol $\xt_{H\vol} \,$ 
in 
(5.1)  is actually 
the equivalence class of Hamiltonian 
vector valued forms modulo an addition of 
pre-Hamiltonian vector valued forms.  
However, the bracket 
\beq
\pbr{H\vol}{\ff{p}}'=\frac{1}{p!} 
\tilde{X}^v_{\cdot k} \ dx^k \wedge 
\der_v F\lind{i}{p} dx^{i_1}\we ... \we dx^{i_p} 
\eeq 
in general cannot be understood as given in terms of the 
equivalence class 
\linebreak{}  
\mbox{$[\tilde{X}$$_{H\vol}\quad $mod$\tilde{X}{}_0]$}, 
as it obviously depends on the choice of a particular 
represenative $\tilde{X}$$_{H\vol}$ 
(this is in particular related 
to the lack of the antisymmetry of the bracket). 
Nevertheless for some forms, 
at least those which belong to the center 
of generalized Gerstenhaber algebra of graded Poisson brackets of forms, 
this dependence cancels out 
so that the bracket can be evaluated ``off-shell''. 
The complete characterization of forms for which the bracket 
(5.8) does not depend on the choice of the representative 
$\tilde{X}$$_{H\vol}$ and for which the bracket (5.8) is graded 
antisymmetric is as yet not studied.  
For other forms the bracket with $H\vol$ can only be understood 
``on-shell'' in the sense specified below. 

Namely, 
let us note 
that $\tilde{X}$$_{H\vol}$ 
may be regarded,  
along with the canonical $n$-vector field $\xx{n}$ in Sect. 2, 
as a suitable analogue 
of the canonical Hamiltonian vector field whose integral curves 
are known to be the trajectories of a mechanical system in the phase space. 
Indeed, the expression of the components of  $\tilde{X}$$_{H\vol}$ 
in (5.7) gives rise to the DW Hamiltonian field equations (2.3) 
if the geometrical object 
associated with the vector valued form 
$\tilde{X}$$_{H\vol}$ 
and generalizing the integral curves of a vector field
is given by  
\beq
\tilde{X}^v_{\cdot k} = \frac{\der z^v}{\der x^k}.       
\eeq
Geometrically, eqs. (5.9) describe a web of $n$ integral curves 
associated with  the field of vector valued forms.      
This  web spans an  $n$-dimensional   surface in the polymomentum 
phase  space 
	and  
represents a solution of 
field equations.    
The bracket in (5.8) is said to be evaluated ``on-shell'' if the 
value of $\xt^v{}_k$ in the right hand side of (5.8) 
is taken ``on-shell'', 
i.e. on a specific section $z^v=z^v(x)$,  
on which it is given by (5.9). 

Now, let us define the {\em total exterior differential}, $\bd$,   
of a $p$-form: 
\beq 
\bd \ff{p}
:= \frac{1}{p!}\der_v F\lind{i}{p} (z^v,x^i)\frac{\der z^v}{\der x^i} 
dx^i \we dx^{i_1}\we ... \we dx^{i_p}  
+ d^{hor}\ff{p} ,     
\eeq 
where the latter term  
$$ d^{hor}\ff{p}:=  
\frac{1}{p!}\der_i F\lind{i}{p} (z^v,x^i) 
dx^i \we dx^{i_1}\we ... \we dx^{i_p}. $$  
By comparing (5.10) with (5.8) and (5.9) 
we conclude that  
\beq
\mbox{{\bd}\ff{p}}=\pbr{H\vol}{\ff{p}}'+d^{hor}\ff{p} , 
\eeq 
where the bracket is understood as evaluated ``off-shell'' 
when possible, or   ``on-shell'' otherwise.  
The  equation above can be viewed as the equation of motion 
of a $p$-form dynamical variable in Poisson bracket formulation. 
The left hand side of  (5.11) generalizes the total time derivative 
in the similar equation of motion of a dynamical variable in mechanics: 
$dF/dt=\{ H,F \} + \der F/ \der t$;   
and its last term present if the  dynamical variable explicitly depends 
on space-time coordinates generalizes 
the partial time derivative term. 
It is clear 
also 
that essentially  
${\bd}=\pounds{}_{\tilde{X}^{tot}}$,
where $\tilde{X}^{tot}=\tilde{X}_{H\vol} + \tilde{X}^{hor}\, $ 
and 
$\, \tilde{X}{}_{H\vol}^{hor} := \delta^i_k\,dx^k \otimes \der_i \,$. 
Thus, the equation of motion in the form (5.11) has a similar meaning 
to that of the equation of motion in mechanics: 
it determines a (generalized) Lie-dragging of a dynamical variable 
along the (generalized) Hamiltonian flow. 
In the present context the ``Lie-dragging'' 
with respect to the field of  vector valued forms is  
formally given sense  by the expression for a generalized Lie derivative, 
eq.  (5.6); and the analogue of the ``Hamiltonian flow''  is given by 
eqs. (5.7), (5.9).    

\medskip

It should be noted that the enlarging of the set of Hamiltonian
multivector fields 
by including  the vector valued one-forms 
associated with horizontal 
$n$-forms implies a certain extension of  
graded Lie algebras of Hamiltonian and 
locally Hamiltonian  multivector fields.   
Let us outline some of the problems related to this extension. 
\newcommand{\oldtxtg}{
At first, one should define 
the bracket of two vector valued forms. 
The 
naive 
definition could be 
\beq 
\mbox{\nbr{$\tilde{X}$$_1$}{$\tilde{X}$$_2$} \inn\ $\Omega :=$
$\pounds$$_{\tilde{X}_{1}}$$(\tilde{X}_{2} \inn\ \Omega)$, }
\eeq 
however,  
a formal
degree counting shows that the right hand side is 
identically zero because
\beq 
\mbox{\nbr{$\tilde{X}$$_1$}{$\tilde{X}$$_2$} $\in \D^1_2,$ } 
\eeq 
where $\D^p_q$ 
denotes the space of vertical-$p$-vector valued 
horizontal $q$-forms. 
Nevertheless, eq. (5.12) is exactly a property 
of the Fr\"{o}licher-Nijenhuis bracket
of two vector valued one-forms \cite{FN,Nono}. 
Therefore, it is natural to
identify the bracket of two vertical-vector valued one-forms 
with the (vertical) FN bracket. 
Because of (5.12) and, more generally, 
\beq 
\mbox{\nbr{$\tilde{X}$$_1$}{$\tilde{X}$$_2$}$_{FN}$ 
$\in \D^1_{p+q} , $  } 
\eeq
for $\tilde{X}$$_{1} \in \D^1_p$ and 
$\tilde{X}$$_{2} \in \D^1_{q}$, 
the vector valued forms of higher degrees will also  be involved when 
attempting to close the algebra with respect to the FN bracket. 
However, these objects extend only the space of pre-Hamiltonian
fields (cf. Sect. 3.2), because 
for $\tilde{X} \in \D^1_{p}  $, $p>1$, 
\beq 
\mbox{$\tilde{X} \inn\ \Omega = 0 $, }
\eeq 
where $\inn\ $ denotes the FN inner product of a vertical-vector 
valued $q$-form
$\tilde{X}$ with a form:
\beq 
\tilde{X}\inn\ \mu := 
\frac{1}{q!}
X^{v}_{\cdot k_1 ... k_q} dx^{k_1}\wedge...\wedge 
dx^{k_q} \wedge(\der_{v}\inn\ \mu).
\eeq
}
At first, the bracket of two vector valued forms 
should be specified. The natural choice would be the 
(vertical) Fr\"{o}licher-Nijenhuis (FN) bracket \cite{FN,Nono,kolar} 
which maps the pair of vector valued one-forms 
to a vector valued two-form. 
The result is therefore always vanishing when 
acting on the polysymplectic form. 
Similarly, vector valued forms 
of higher degrees which in principle  
appear when  closing the algebra with respect to the 
FN bracket will also vanish on the polysymplectic 
form, 
thus contributing only to the space of 
pre-Hamiltonian fields.      

Next, a bracket 
of  vector valued one-forms 
with  Hamiltonian multivectors of 
arbitrary degree has to be defined. 
The most naive extension of  
(3.7), 
which we essentially have used when 
defining the bracket  in (5.4),  
is 
\beq 
\nbr{\tilde{X}}{\xx{p}}' \inn\ \Omega:= 
\mbox{$\pounds$}_{\tilde{X}}
(\xx{p} \inn\ \Omega)   ,   
\eeq 
whence it follows that 
$ \nbr{\tilde{X}}{\xx{p}}{}' \in \D^p_1, 
$
where $\D^p_q$ 
denotes the space of vertical-$p$-vector 
valued horizontal $q$-forms. 
In such a manner
the 
multi\-vec\-tor valued 
one-forms 
come into play. 
Note, that the bracket defined in (5.12) 
is marked with the prime 
because it is not graded antisymmetric 
in general.
Further closing of the algebra requires a 
proper definition of the bracket of  multivector valued 
one-forms appearing above. 
To do this would require 
a proper definition of 
a generalized inner product, 
to be also denoted $\inn$, 
by which a multivector valued form
acts on forms. Assuming 
according to  \cite{Vinogr} for $\xt \in \D^p_q$ 
\beq 
\xt \inn\ \Omega := 
\frac{1}{q!p!}
\xt^v{}\uind{i}{p-1}{}\lind{k}{q}
dx^{k_1}\we ...\we dx^{k_q} \we 
\der_v{}_{i_1 ... i_{p-1}} \inn\ \Omega , 
\eeq
it may be concluded that 
the  non-symmetric bracket of 
two multivector valued one-forms,  
defined similarly to (5.12),   
leads to 
multivector valued two-forms and, therefore, 
the algebraic closure of the algebra with respect to this 
kind of bracket  
may involve  multivector valued forms 
$\xt \in \D^p_q$ of all possible degrees $p$ and $q$.

Thus, essentially, we are led to the problem of the construction of  
the bracket operation on arbitrary multivector valued forms, 
which would properly generalize 
the SN bracket on multivector fields 
and the FN bracket on vector valued forms. 
This is related to the problem of  
embedding the SN and FN graded Lie algebras 
into a larger 
algebraic structure on multivector valued forms 
which was recently considered by 
A.M.  Vinogradov \cite{Vinogr}.  
His ``unification theorem'' 
(see also \cite{roger} for a related  discussion)  
states that the SN and FN 
algebras may be embedded in a certain 
{$\Bbb Z$}-graded quotient algebra of
the algebra of ``super-differential operators'', 
or graded endomorphisms,  
on the exterior algebra,  
which are actually represented by  
multivector valued forms.  
 
Note that multivector valued forms  
essentially have already 
appeared in Sect. 3.1 
when constructing the  analogue of the multivector 
field associated with the exterior product of two Hamiltonian forms 
(see eq. 3.25). This suggests that multivector valued forms 
can be associated with 
non-Hamiltonian horizontal forms 
and in this way the graded Poisson bracket can be 
extended to the latter.   
An extension of this sort is postponed to a future publication 
(see also \cite{frontiers} for a preliminary discussion).

{\footnotesize \sl Note added in the final revision (July 1997)}.  
{\footnotesize After the paper was completed a further progress 
has been made in extending the bracket to   
non-Hamiltonian  forms. This extension leads to a non-commutative 
(in the sense of Loday) graded Lie algebra structure  
(see \cite{kan-goslar,kan-bial}) and the corresponding 
non-commutative generalization of higher-order and right Gerstenhaber 
algebras mentioned in Sect. 3. Furthermore, 
the product operation preserving the space of Hamiltonian forms is 
presented in  \cite{kan-bial} which turns the latter 
into a  
 Gerstenhaber algebra. 
}

\section{\normalsize\bf Several simple applications}

\subsection{\normalsize\bf Interacting scalar fields}

As the simplest example of how the formalism  constructed in the 
previous sections works,
 consider a system of interacting real scalar fields
$\{\phi^{a}\}$ which is described by the Lagrangian density
\beq
L=  \frac{1}{2}\der_i \phi^a \der^i \phi_a - V(\phi^a).
\eeq
Henceforth the space-time is assumed to be Minkowskian.
The polymomenta derived from (6.1) are
\beq
p^i_a := \frac{\der L}{\der(\der_i \phi^a)} =  \der^i \phi_a
\eeq
and for the DW Hamiltonian function we easily obtain
\beq
H=  \frac{1}{2}p^i_a p^a_i + V(\phi).
\eeq 
In terms of the canonically conjugate (in the sense of Sect. 4.1)
variables 
\[ \phi^a \quad {\rm and} \quad \pi_a := p^i_a \dvol{i}\] 
which have the
following non-vanishing  Poisson bracket
\beq
\pbr{\phi^a}{\pi_b}  =-\delta^a_b, 
\eeq 
we can also write
\beq
H\vol =  \frac{1}{2} (\star \pi^a) \wedge \pi_a + V(\phi)\vol.
\eeq
where $\star \pi_a :=  p_a^i dx_i$.
Now, the canonical DW field equations may be written 
in terms of the Poisson brackets on forms: 
\begin{eqnarray}
\mbox{{\bd }}\pi_a  &=\, \pbr{H\vol}{\pi_a}&=
\, \xt_{H\vol}\inn d\pi_ a   \nn \\
& &=\; \xt^k_{ak}\vol      \nn \\
& &=\; -\der_a H\vol,  \\
                             &                    &        \nn       \\
\mbox{{\bd}}\phi^a &=\, \pbr{H\vol}{y^a}  &=\,
\xt_{H\vol}\inn dy^a  \nn \\
& &=\; \xt^a{}_k dx^k  \nn \\
& &=\; \der^a_k H dx^k  \nn \\
& &=\; p^a_k dx^k       \nn \\
& &=\; \star    \pi^a ,  
\end{eqnarray} 
It is easy to show that eqs. (6.6), (6.7)   are equivalent to the 
field equations following from the Lagrangian (6.1):
$$\Box \phi_a = -\der_a V. 
$$ 

\subsection{\normalsize\bf The electromagnetic field}

Let us start from the conventional Lagrangian density 
\beq  
L = -\frac{1}{4}F_{ij}F^{ij} - j_{i}A^{i},
\eeq
where $F_{ij}:=\der_i A_j - \der_j A_{i}$. For the polymomenta
we obtain
\beq
\pi^{i}_{\cdot m}:= \frac{\der L}{\der(\der_i A^m)}= -F^{i}_{\cdot m}.
\eeq
whence
the "primary constraints"
\[ \pi^{im}+\pi^{mi}=0 \]
follow. 
The DW Legendre transformation 
$\der_i A^m \rightarrow \pi^i_{\cdot m}$ 
is singular,   
so that the use of the naive  
DW Hamiltonian function 
\beq
H= -\frac{1}{4}\pi_{im} \pi^{im} + j_{m}A^{m} 
\eeq
 leads to the incorrect equation of 
motion 
$\der_{i}A^{m}=\der H / \der \pi^{i}_{\cdot m} 
= 1/2 
\pi_{i}^{\cdot m}$.  
The problems of this
sort are usually handled by adding the constraints with some Lagrange
multiplies to the canonical Hamiltonian function 
and then applying the  well known 
Dirac's procedure for constrained systems. 
This approach  could in principle be extended to 
singular, from the point of view of DW theory,  Lagrangean theories. 
However, our formalism provides us with  
another possibility which is  
based on the freedom in choosing 
the canonically conjugate field 
and momentum variables 
which was   mentioned in Sect.  4.1.  

    Namely, let us  take as the 
canonical field variable the
one-form 
\[ \alpha:=A_{m}dx^{m}\] 
per se,  instead of the set of its components
$\{A_{m} \}$ considered usually. 
Then, the momentum variable canonically conjugate to $\alpha$ 
may be chosen to be the $(n-2)$-form 
\[ \pi := -F^{im}\, \der_i \inn \der_m\inn \vol=\pi^{im}\der_i \inn 
\der_m \inn \vol   \] 
in which essentially the dual of the Faraday 2-form $1/2 F_{ij}dx^i\we dx^j$ 
can be recognized. 
This is evident from the following calculation of the Poisson bracket 
of the 1-form $\alpha$ with the $(n-2)$-form $\pi$.

The components of the bivector field $X_{\pi}:=X^{vi}\der_v\we\der_i$
associated with $\pi$  are given  by
\beq
X_{\pi} \inn \Omega = d\pi =
d \pi^{im}\we \der_i \inn 
\der_m \inn \vol   ,
\eeq 
where the polysymplectic form $\Omega$ is 
\[\Omega = -dA^m \wedge d\pi^{ i}{}_{m} \wedge \dvol{i}.  \]
The only non-vanishing components of $X_{\pi}$ are obviously 
those along the field 
directions $\der_{A_m}$, 
so that  
$X_{\pi}=X^{A_m i}\der_{A_m}\we \der_i$.
Substituting this expression to 
the left hand side of 
(6.11) we obtain 
\beqa
X_{\pi}\inn\Omega &=& - X^{A_m i}[\der_{A_m}\otimes\der_i 
- \der_i \otimes \der_{A_m} ] \inn 
 dA^n \wedge d\pi^{ j}{}_{n} \wedge (\dvol{j}) \nn \\
&=& 2 X^{A_m i} [\der_i \otimes \der_{A_m} ] \inn 
dA^n \wedge d\pi^{ j}{}_{n} \wedge (\dvol{j}) \nn \\
&=&- 2 X^{A_m i} d\pi^j{}_m \we \der_i \inn \der_j \inn \vol .
 \nn 
\eeqa
A comparison with the  right hand side 
of (6.11) 
gives
\beq
 X^{A_m i} = -\frac{1}{2} g^{mi} ,
\eeq
where $g^{ik}$ is the metric tensor on the space-time. 
Thus, 
\beq
\pbr{\pi}{\alpha}=\xx{2}_{\pi}\inn d\alpha
=-2X^{A_m i} (\der_i\otimes \der_{A_m}) 
\inn (dA_n\we dx^n) = n .
\eeq

Eq. (6.13) justifies our choice of the $(n-2)$-form $\pi$ 
as the canonically conjugate variable 
to the one-form potential $\alpha$, as the bracket reduces to 
the canonical bracket when $n=1$.      
Note that 
the form $\pi$ is a 
Hamiltonian form, in contrast to its dual, 
the Faraday 2-form,  
which we might naively try to associate 
with $\alpha$  as its conjugate momentum.  

In terms of the variables $\alpha$ and $\pi$ 
the DW Hamiltonian $n$-form can be written as follows 
\beq
H\vol = 
\frac{1}{2}	
\star \pi \we \pi  + \alpha \wedge j,
\eeq
where $j:= j^i \dvol{i}$  is the electric current density $(n-1)$-form. 
The Maxwell equations may now be written 
in 
DW Hamiltonian form in
terms of variables $(\alpha, \pi)$ and the 
graded Poisson bracket of forms:
\begin{eqnarray}
\mbox{{\bd}}\alpha =\pbr{H\vol}{\alpha}&=&\star^{-1}\pi,
  \nonumber \\
                                                            & &      \\
\mbox{{\bd}}\pi = \pbr{H\vol}{\pi}     &=&j. \nonumber
\end{eqnarray}
Thus, 
the DW Hamiltonian formulation of  
Maxwell's electrodynamics is obtained 
without any recourse to the formalism
of 
  the 
  fields with constraints. 
The constraints, 
however, 
both gauge and initial data,  
did not disappear, 
of course, but they 
	can be taken into account 
after the covariant DW Hamiltonian formulation 
	is constructed.
Note that in general the systems which 
have Hamiltonian constraints in the sense 
of the usual instantaneous Hamiltonian formalism 
have a totally different structure of constraints 
(understood as the obstacles to the covariant Legendre transform) 
when viewed from the perspective of the 
DW Hamiltonian formulation, 
or even 
can appear to  be constraint free 
within the latter, 
as the following example illustrates.

\subsection{\normalsize\bf The Nambu-Goto string}

\newcommand{\xdot}{\stackrel{.}{x}}
\newcommand{\xprim}{x'}

The classical dynamics of a string sweeping in space-time the world-sheet 
$x^a = x^a (\sigma,\tau)$ is given by the Nambu-Goto Lagrangian
\beq    
L= -T\sqrt{(\xdot \cdot \xprim)^2 - \xprim^2 \mbox{$\xdot$$^2$}} = 
-T\sqrt{-det\|\der_i x^a \der_j x_a \|}, 
\eeq
where $\xdot^a := \der_{\tau} x^a, \xprim^a := \der_{\sigma} x^a$, 
$T$ is the 
string rest tension, 
and the following notation for the 
world-sheet parameters 
$(\sigma,\tau) = (\tau^0 , \tau^1):=(\tau^i); \; i=0,1$ is used. 

  Let us define the  polymomenta:
\begin{eqnarray}
p^0_a := \frac{\der L}{\der \mbox{$\xdot$$^a$}}  = T^2\frac{(\xprim 
 \cdot \xdot)\xprim_a - \xprim^2 \xdot_a}{L},& & \nonumber \\
      &                                 &           \\
p^1_a := \frac{\der L}{\der \xprim^a} = T^2\frac{(\xprim 
 \cdot \xdot)\xdot_a - \mbox{$\xdot$$^2$} \xprim_a}{L}.& &   \nonumber 
\end{eqnarray} 
From eqs. (6.17) the following identities follow
\begin{eqnarray}
p^0_a \xprim^{a}=&0, \hspace*{4em} (p^0)^2 + T^2 \xprim^2&= 0, \nonumber \\
                 &                                       &    \\
p^1_a \xdot^{a} =&0, \hspace*{4em} \mbox{$(p^1)^2 + T^2 \xdot$$^2$}&= 0.
 \nonumber
\end{eqnarray}
However, these identities  do  {\em not} have the meaning of Hamiltonian
constraints within the DW Hamiltonian formalism,  
as they do not imply
any relations between the  coordinates $x^a$ and the
polymomenta $p^i_a$. 
In fact, 
eqs. (6.17) can be easily solved (if $L \neq 0$) 
yielding expressions for 
the generalized velocities $(\xdot, \xprim)$ 
in terms of the polymomenta; 
this proves {\em de facto} that the DW Legendre transform
$( \xdot , \xprim ) \rightarrow ( p^0_a, p^1_a )$ 
for the Nambu-Goto Lagrangian is regular. 

      In terms of the polymomenta the DW Hamiltonian function 
takes the form 
\beq
H=-\frac{1}{T}\sqrt{-det\|p^i_a p^{aj}\|}   ,
\eeq
and can also be expressed in terms of the one-form momentum variables 
\[\pi_a := p^i_a \varepsilon_{ij}d\tau^j , \] 
which are canonically conjugate 
(in the sense of Sect. 4.1) to $x^a$.  
In fact, 
\[det\|p^i_a p^{aj}\|=\frac{1}{2}(\varepsilon_{ij}p^i_a p^j_b)
(\varepsilon_{ij}p^{ai}p^{bj}) , \]
and
\[\varepsilon_{ij}p^i_a p^j_b =  \star^{-1} (\pi_a \wedge \pi_b).  \] 
The string equations of motion in terms of the graded Poisson 
brackets of forms can now be written as follows 
\begin{eqnarray} 
\mbox{{\bd}$x^a  $}&=\;  \pbr{H\vol}{x^a}\; = 
             &\frac{\der H}{\der p^i_a}d\tau^i, \nonumber \\ 
             & &   \\
\mbox{{\bd}$\pi_a$}&=\; \pbr{H\vol}{\pi_a}\; =&0. \nonumber
\end{eqnarray} 

As yet another application 
let us  show that the Poincar\'{e} algebra 
corresponding to the internal symmetry of the $x$-space in string theory 
can be realized with the help of our brackets on forms. 
In the $x$-space the translations 
and the Lorentz rotations  
are generated 
by the vector field $X_a := \der_a$ 
and  the bivector field $X_{ab} := x_a \der_b - x_b \der_a$ 
respectively. 
The corresponding conserved current densities are given by 
one-forms:
\beq 
\mbox{$\pi_a$$  \;\;$ {\rm and} 
$\;\; $$\, \mu_{ab}:=x_a \pi_b - x_b \pi_a . $ }
\eeq
One can easily check that the conservation laws 
\beq 
\mbox{\bd} \pi_a = 0,   
\quad  \mbox{\bd}
\mu_{ab} = 0 
\eeq
follow from the string equations of motion.
Now, a straightforward calculation of the Poisson brackets of 
one-forms  
above  
yields:
\begin{eqnarray}
\hspace*{2em}\pbr{\pi_a}{\pi_b}&=&0, \nonumber  \\
\pbr{\mu_{ab}}{\pi_c}          &=&g_{ac}\pi_b - g_{bc}\pi_a,   \\
\pbr{\mu_{ab}}{\mu_{cd}}       &=&C_{abcd}^{ef} \mu_{ef} \nonumber,        
\end{eqnarray}
where $g_{ab}$ is the $x$-space metric and
\[C_{abcd}^{ef}=-g^e_cg^f_ag_{bd} + g^e_cg^f_bg_{ad} - 
g^e_ag^f_dg_{bc}  + g^e_bg^f_dg_{ac} \] 
are the Lorentz group structure
constants. Thus the internal Poincar\'{e} symmetry of a string is
realized in terms  of the Poisson brackets of one-forms which 
correspond  to the conserved currents related to this symmetry.


\section{\normalsize\bf Discussion} 

The subject of the present paper is 
an extension  of basic   
structures of the mathematical formalism 
of classical Hamiltonian mechanics to 
field theory within the De Donder-Weyl  
polymomentum Hamiltonian formulation. 
As a subsequent application to the problem of quantization of field theories 
is implied,  emphasis is given to the analogues of the structures 
and constructions which are important for different quantization 
procedures such as canonical, geometric or deformation.  

Our starting point 
is the Poincar\'e--Cartan 
$n$-form 
corresponding to the 
De Donder-Weyl theory 
of  multiple integral variational problems.  
We show that the  
De Donder-Weyl 
Hamiltonian  field equations 
can be formulated in terms of  
the multivector field of degree $n$, eq. (2.5),  
which annihilates the exterior differential of the 
 Poincar\'e--Cartan form, 
and whose integral $n$-surfaces represent solutions of 
field  equations. 
This observation leads us to the notion of 
the polysymplectic form of degree $(n+1)$. 
The polysymplectic form is defined 
in local coordinate terms 
as the vertical exterior differential  
of the non-horizontal part, $\Theta^V$,  of the 
Poincar\'e-Cartan form, see eq. (3.1). 
This definition, however,  essentially implies 
a triviality of the extended polymomentum phase space 
as a bundle over the space-time manifold 
unless 
the potential of the polysymplectic form, $\Theta^V$, 
the vertical exterior differential,  $\dv$, 
and  the polysymplectic form itself 
are understood as appropriate cosets 
(see also \cite{kan-goslar}, cf. \cite{crampin}). 
The same concerns the equations including these objects.  

Obviously, it would be highly desirable to reveal an intrinsically 
geometric formulation which could reproduce the essential features 
of the construction of the present paper. 
This problem, however, requires more involved 
intrinsic geometric techniques and the jet bundle language 
(see e.g. \cite{Sardan,crampin,echev1,echev2,deleon,saunders})  
using 
of 
which we  avoided  here.  
   
\newcommand{\txda}{  
Note, however, that in a more general case the potential $\Theta^V$ 
of the polysymplectic form (see Sect. 3) 
can in principle be understood as a representative of the 
equivalence class of the PC forms modulo horizontal $n$-forms 
\cite{kan-goslar} (see also \cite{crampin}), 
and the vertical 
differential  $\dv$ can be understood as the usual exterior differential 
modulo the horizontal exterior differential. Vertical $p$-multivectors 
can also be given the intrinsic meaning as the $p$-multivectors vanishing 
on arbitrary horizontal $p$-form, which means that they have 
{\em at least} one vertical component. 
}

Within the present approach,  
the polysymplectic form 
plays 
the 
role similar to that of the symplectic form in mechanics
to which it reduces when $n=1$.  
Unlike the latter the polysymplectic form is not
purely vertical - 
it has two vertical and $(n-1)$ horizontal 
components - 
and it   
is also not non-degenerate in the sense that 
a certain class of  
multivector fields called pre-Hamiltonian exists 
on which it vanishes.  
Note that the construction of the present paper 
could be also carried out with only minor  modifications 
for  possible alternatives of the polysymplectic form: 
$dy^a\we dp_a^i \otimes \der_i \inn \vol$ (see  \cite{Kij+Tul}), 
$dy^a\we dp_a^i \otimes \der_i$  (see \cite{Guenther87a})   
and $dy^a\we dp_a^i \we \vol \otimes \der_i$ (see \cite{Sardan}). 
Our main reason for prefering the polysymplectic form 
(3.1) 
is that it comes into being immediately from the 
Poincar\'e-Cartan form. 
This relationship  
may allow us  to guess 
the analogues of the polysymplectic form in more general 
Lepagean canonical theories for fields 
(see e.g. \cite{Kastrup83,giaq}), 
such as the Carath\'eodory theory \cite{Carath29,Rund}, for instance.

The basic symmetry of the theory, 
that of the polysymplectic form,  
is formulated as a statement of the vanishing of 
the generalized Lie derivatives 
of the polysymplectic form with respect to the vertical multivector 
fields of degrees $1\leq p \leq n $, called locally Hamiltonian; 
see eq. (3.9).  
This graded symmetry may be 
viewed as  a field-theoretical extension of the canonical symmetry 
known in mechanics. 
We show that the set of locally Hamiltonian multivector 
fields is a graded Lie algebra with respect to the 
(vertical) Schouten-Nijenhuis 
bracket     
which is 
defined with the help of the 
generalized Lie derivative with respect to a vertical multivector field.

Furthermore,  
the polysymplectic form 
gives rise to 
the  map between  vertical multivector fields 
and horizontal forms, see eq. (3.10). Horizontal forms  
play the role of dynamical variables in the present formalism. 
The map (3.10) generalizes the map between  
dynamical variables
and  Hamiltonian vector fields  which is given in mechanics 
by the symplectic form. 
Thus we are  led to the  notions of a Hamiltonian 
multivector field, to which the horizontal form can be associated, 
and a Hamiltonian form, to which the vertical multivector field can be 
associated. 
Hamiltonian multivector fields also form a graded Lie 
algebra with respect to the vertical Schouten-Nijenhuis bracket. 
This algebra is in fact an ideal  
in the graded Lie algebra of locally Hamiltonian fields. 
 
The existence of the multivector field associated with a horizontal 
form imposes certain restrictions on the  dependence of 
the latter on the polymomentum variables. 
These restrictions
specify the 
class of admissible horizontal forms called Hamiltonian. 
For instance, 
in the case of forms of degree $(n-1)$ 
only the specific linear dependence on 
polymomenta is admissible, see eq. (4.5). 
As the result, the  space of Hamiltonian forms is not 
stable under the exterior product. 
This poses the question as to how (or whether)  
the 	
construction of the present paper 
can be extended to more general horizontal forms, 
or whether there exists a proper product operation on the 
space of Hamiltonian forms (cf. \cite{kan-bial}).     
 
It should be noted  that the multivector field 
associated with a Hamiltonian form  
is not generally uniquely specified.   
The arbitrariness is related to the existence of 
pre-Hamiltonian fields  which annihilate the polysymplectic form.
However, as the pre-Hamiltonian fields form an ideal 
in the graded Lie algebra of Hamiltonian 
multivector fields,  
 the map in eq. (3.10) from forms to multivector fields is  
essentially 
a map from forms to the {\em equivalence classes} of 
Hamiltonian multivector fields modulo 
an addition of pre-Hamiltonian fields.

Further, the vertical Schouten-Nijenhuis 
bracket of Hamiltonian multivector 
fields induces 
a bracket operation on 
Hamiltonian forms.   
This is our graded Poisson bracket operation on forms introduced 
in Sect. 3.  
It equips  the space of 
Hamiltonian forms with the structure of 
{$\Bbb Z$}-graded Lie algebra. 
There are also two ways in which this bracket operation 
generalizes the known derivation property  
of the usual Poisson bracket.  
The first leads to the notion of a 
{\em higher-order} 
Gerstenhaber algebra with respect to the operations 
of the exterior product and the Poisson bracket on forms. 
This means that the 
graded Leibniz rule in the definition of a Gerstenhaber algebra is 
replaced by a more involved expression which has a natural interpretation 
of a higher-order Leibniz rule. The formulation in terms of the 
$\Phi$-maps introduced by Koszul in his discussion of graded higher-order 
differential operators on graded commutative algebra \cite{koszul} 
proved to be useful here.  
The second way, which is based on the fulfillment of  
the right graded Leibniz rule, eq. (3.26),  leads to the 
right Gerstenhaber algebra structure. 
It should be noted, however, 
that the considerations of the analogues
of  graded Leibniz rules, both left and right, 
with respect to  the exterior product of forms, 
are not sufficiently well-grounded 
because the graded Poisson bracket is  
defined here only on the subspace of Hamiltonian forms 
which is  not stable under the exterior product. 
Correct consideration would require  
an extension 
of the definition of graded Poisson bracket 
to arbitrary horizontal forms 
which we hope to discuss in a forthcoming paper 
(see also \cite{frontiers}); 
cf. \cite{kan-goslar,kan-bial} for a further progress.     

Note in passing, that our 
graded Poisson bracket on exterior forms  
is different from other  
bracket operations 
which can be  defined on forms by means of the Poisson bivector 
(see e.g. \cite{koszul,Vinogr,ksmagri,karas,michor})  
in the context more close to that of mechanics than to field theory.  
An interesting graded extension of the Poisson bracket 
is also constructed in \cite{ibort}. 

As an application of our 
graded Poisson bracket of forms, 
the Poisson bracket formulation 
of the equations of motion of dynamical 
variables represented by horizontal forms 
is considered.  
This form of the equations of motion 
is given by the Poisson bracket
with the DW Hamiltonian function, 
as it may be  anticipated  
by  analogy with the canonical formalism in mechanics. 
More precisely,
the Poisson bracket of 
a Hamiltonian $(n-1)$-form with the DW Hamiltonian function  $H$
generates the inverse Hodge dual of the total 
(i.e. taken on sections) 
exterior differential of the form, see eq. (4.8). 
The DW Hamiltonian form of 
field equations results from 
this more general statement, when the suitable $(n-1)$-forms 
linearly constructed from the polymomenta or the field variables 
are substituted 
for a Hamiltonian $(n-1)$-form into the bracket, see eq. (4.9).

A generalization of 
the Poisson bracket formulation of the equations 
of motion to  forms of arbitrary degree requires a certain 
extension of the construction outlined above.
Namely, the space of Hamiltonian forms has to be enlarged 
by adding  horizontal forms of degree $n$ and, 
correspondingly, 
the space of Hamiltonian multivector fields 
by adding  
objects of formal degree zero, the vertical-vector valued 
horizontal one-forms which are associated with $n$-forms.
This allows us to define the 
(left) 
bracket operation with $n$-forms and to show 
that the bracket of a form 
with the DW Hamiltonian $n$-form, $H\vol$, 
generates 
the total exterior differential (see  the definition in Sect. 5)
of the former, i.e. 
\mbox{{\bd}$ \, \cdot \, = 
\pbr{H\vol}{\, \cdot \, }'$.} 

The appearance of  vector valued one-forms 
associated with $n$-forms  
enlarges the space of Hamiltonian multivector fields  
	and implies a  certain extension
of the corresponding graded Lie algebra. 
	The algebraic closure of this extension 
seems to involve multivector valued   
forms of higher degrees 
and may require  
an  appropriate  definition
of a bracket operation on these objects  
which would generalize 
the 
Lie, Schouten-Nijenhuis (SN) 
and Fr\"{o}licher-Nijenhuis (FN) brackets. 
The latter  
problem is related to the  construction of an  
algebraic structure on multivector valued forms 
which  unifies  the SN and FN graded 
Lie algebras of, respectively,  
multivector fields and vector valued forms.      
In this connection the results of A.M. Vinogradov \cite{Vinogr} 
are of great interest.     

It is natural to expect  that 
the consideration of all the
elements of 
the hypothetical bracket algebra of multivector valued forms 
may  allow 
us to  avoid the restrictive conditions on 
the 
admissible dependence of Hamiltonian forms on polymomenta. 
In other words, the expectation is that 
objects of more general nature than multivectors 
namely, the multivector valued forms, 
can  be associated 
with non-Hamiltonian  
horizontal forms, 
thus opening a possibility 
of extending  the construction 
of the present paper to arbitrary horizontal forms.   
A preliminary  discussion of this possibility can be found in 
\cite{frontiers} 
(see also \cite{kan-goslar,kan-bial} 
for a subsequent development).   

Note, 
that there are several reasons 
mentioned in the text which make 
an extension of the construction of 
graded Poisson bracket  beyond the space 
of Hamiltonian forms desirable and necessary. 
First is that only in this case the generalized 
graded Poisson properties
of the bracket, eqs. (3.22) and (3.27), 
may be substantiated. 
Second is that 
there exist dynamical variables in field theory which cannot 
be naturally related to Hamiltonian forms. 
The energy-momentum tensor mentioned 
in Sect. 3.1 is an example.  
Besides, the examples considered in Sect. 6 
show  that the Hodge duals of Hamiltonian forms 
(as, for instance, the one-form $\star \pi^a$ in Sect. 6.1) 
naturally appear in the DW formulation of particular models, 
whereas  it is not difficult to see that if a 
Hamiltonian form depends on 
polymomenta then its Hodge dual is not a 
Hamiltonian form. 

\medskip 

Let us now sketch the connection of the formalism 
constructed in this paper with the conventional instantaneous 
Hamiltonian formalism for 
fields (cf. \cite{sniat,Gotay multi2,Gotay ea}).    
Let us choose a
space-like surface $\Sigma$ in the $x$-space (here we will assume the
latter to
be pseudo-euclidean  with  the signature $++...+-$). The restrictions of
the polymomentum phase space variables  to $\Sigma$ will be  functions of
the $x$'s. In particular, if $\Sigma$ is given by the equation $x^n = t$
($n$ denotes the number of the time-like component of
$\{x^i\}=\{x^1,...,x^{n-1},x^n\}$, and  is not an index), we have
$y^a|_{\Sigma}=y^a({\bf x},t)$ and $p^i_a|_{\Sigma}=p^i_a({\bf x},t)$,
where {\bf x} denotes the space-like components of $\{x^i\}$. Moreover,
the restriction of forms to $\Sigma$ implies that we must set 
$dx^n=0$, so that
for $p_a:=p^i_a\der_i\inn \vol$ we obtain $p_a|_{\Sigma}=p^n_a({\bf
x},t)\der_n\inn \vol$, where $\der_n \inn \vol$ is clearly the
$(n-1)$-volume form on $\Sigma$, which we shall denote as $d\bf x$. 
The functional symplectic 2-form $\omega$ on the phase space of the
instantaneous formalism may now be related  to the restriction of the
polysymplectic form $\Omega$ to $\Sigma$ in the following way (cf.
\cite{Gotay multi2,Gotay ea}):
\beq
\omega=\int_{\Sigma}(\Omega|_{\Sigma})=
-\int_{\Sigma} dy^a({\bf x}) \wedge dp^n_a({\bf x}) d{\bf x}. 
\eeq
In addition,
the equal-time Poisson bracket  of $y^a({\bf x})$ with the 
canonically
conjugate momentum $p^n_a({\bf x})$:  
\beq
\{ y^a({\bf x}),p^n_b({\bf y})\}_{PB}=
- \delta^a_b\delta({\bf x}-{\bf y}) 
\eeq
may be related to the Poisson bracket of the canonically conjugate
variables $y^a$ and $p_a$ of the DW theory (see Sect. 4.3) as follows:
\beq
\int_{\Sigma_x}\int_{\Sigma_y}\{y^a({\bf x}), p^n_b ({\bf y})\}_{PB} 
f({\bf x}) g({\bf y}) d{\bf x}d{\bf y} = 
\int_{\Sigma} \pbr{y^a}{p_b}    
f({\bf x}) g({\bf x})d{\bf x},  
\eeq
where $f({\bf x})$ and $g({\bf x})$ 
are  test functions. 
In general, one might assume that 
the following relationship between the generalized Poisson bracket of
Hamiltonian forms and the equal-time Poisson bracket of their
restrictions to the space-like surface $\Sigma$ will hold
\beq
\int_{\Sigma_x}\int_{\Sigma_y} 
\{ (\phi_1  \wedge \ff{p}_{1})|_{\Sigma_x}(x), 
(\ff{q}_2 \we \phi_2 )|_{\Sigma_y}(y) \}_{PB} 
 \sim\, \int_{\Sigma} \phi_1 \wedge 
\pbr{\ff{p}_1}{\ff{q}_2}\wedge \phi_2, 
\eeq
where $\phi_1$ and $\phi_2$ denote  horizontal "test forms" 
of degree $(n-p-1)$ and $(n-q-1)$ respectively 
with components depending on the space-time coordinates only, 
 and where the standard 
Poisson bracket $\{\;,\;\}_{PB}$ of forms is defined 
via the standard Poisson brackets of their components 
which survive after the restriction to a Cauchy surface.  
The above formula 
reproduces, in particular, the canonical
equal-time Poisson brackets from the  Poisson brackets of
forms corresponding to 
the  pairs of canonically conjugate variables 
(see Sect. 4.3).  However, 
it does not enable us to  reproduce 
all 
Poisson brackets of interest  in field theory,
because some of such brackets 
involve quantities, an example being the energy-momentum tensor,  
which cannot naturally be related to  the  forms belonging to 
the restricted class of  Hamiltonian forms. 
Furthermore, the brackets 
of interest in field theory may involve higher 
(space-like) derivatives  
of field variables and it is not clear how such dynamical 
variables can be accounted for in the context of  DW theory. 
Both of these are additional indications that 
the approach of this paper  
should be extended to 
include  more general objects than Hamiltonian forms. 

\medskip
 
We would also like to pay attention to the fact that 
the algebraic structures
that arose  in the present 
formulation of classical field theory 
are cognate with those which appeared in the
BRST approach to field theory, in particular, in the
antibracket formalism (see, for example,
\cite{Henneaux book}). 
The latter are, of course, 
established within the framework which is conceptually different 
from the spirit of this paper. 
Nevertheless, a deeper relationship 
than a simple algebraic analogy may not be unexpected 
in view of the above connection between the usual 
Poisson bracket and 
the graded Poisson bracket  of  forms put forward 
here; 
in principle, this might shed light on the geometrical 
origin of the structures of the BRST and BV formalism.  
For a related discussion of interest in this context 
see also \cite{henn}. 
It is also   
worth  noting 
that the structure of 
Gerstenhaber algebra whose generalizations  
arose in our construction 
of the  graded Poisson bracket on forms 
has appeared recently 
in the context  of the BRST-algebraic 
structure of string theory
\cite{LiZuck93}.

\medskip

Note in conclusion  that the polysymplectic form 
and its analogues 
in more general Lepagean canonical theories 
than  the De Donder-Weyl theory considered here 
can be used to define 
the $n$-ary operations of the Nambu bracket type 
in field theory  (see \cite{kan-bial}).    
This opens yet another possibility 
of generalizing the canonical formalism to field theory 
using 
an $n$-ary bracket operation of scalar quantities 
instead of 
the  binary bracket operation on differential forms 
introduced here.  

\medskip

The most intriguing  question for further 
research is 
whether an appropriate quantization 
of graded Poisson brackets  of forms 
can lead to a 
new, inherently covariant   
formulation of 
the quantum theory of fields 
based on the polymomentum formalism.    
This poses, in particular,  
an interesting mathematical problem of quantization, 
or deformation, of a generalized 
Gerstenhaber algebra  
	of graded Poisson brackets of forms  
(or its appropriate subalgebra,  
for the limitations of an analogue of the 
van Howe--Groenewold no-go theorem 
(see e.g. \cite{emch}) 
will certainly take place here).   
A subsequent problem would be the physical 
interpretation of the resulting quantum theory 
and the clarification of its possible relation  
to the well-established part 
of modern  quantum field theory.    

\bigskip

\catcode `\@=11
  
\def\theequation{A.\arabic{equation}} 
   
\catcode `\@=12

{\normalsize\bf Appendix: Higher-order graded Leibniz rule} 

\bigskip

\noindent 
Here we  consider the analogue of the left Leibniz rule for the graded 
Poisson bracket. The expression
\beqa
\label{ad1}
\pbr{\ff{n-p}}{\ff{q}\we \ff{r}}&=&(-1)^p
\xx{p}\inn\dv (\ff{q}\we \ff{r}) \nn \\
&=&(-1)^p\xx{p}\inn ( \dv \ff{q}\we \ff{r} 
+ (-1)^q \ff{q}\we \dv \ff{r} )      
\eeqa
has to be calculated. 
Here the vertical multivector field $\xx{p}$ 
acting via the inner product on the exterior forms 
may be viewed as
a graded differential operator on exterior forms which is composed 
of one vertical derivation of degree $-1$ which is represented by 
the operation $\der_v\inn$,
and  $(p-1)$ horizontal derivations of degree $-1$ 
represented by  operations 
$\der_i \inn$. In fact, 
\beqa
\xx{p}\inn &=& 
\frac{1}{p!}
X^v{}\uind{i}{p-1}\der_v\we \der_{i_1} \we ...
\we \der_{i_{p-1}} \inn \nn \\
&=&(-1)^{p-1}p X^v{}\uind{i}{p-1}
\der_{i_1} \otimes ...
\otimes \der_{i_{p-1}} \otimes \der_v \inn .
\eeqa
It can be said that 
the differential operator on exterior algebra is of order $p$
if it can be represented as a composition  
of $p$ "elementary" graded derivations. 
The latter are classified by
the Fr\"olicher-Nijenhuis theorem \cite{FN} 
and can be of the inner product ($i_*$) or the differential ($d_*$) type. 
The derivations of the inner product type are given 
by the inner products with vector fields - graded derivations 
of degree $-1$ - 
or by the  Fr\"olicher-Nijenhuis inner product with 
vector valued forms - graded derivations of degree $q\geq 0$.
Note that the present  notion of the 
higher-order graded differential operators 
is consistent with the more general definition given by 
Koszul \cite{koszul}.
It is clear now, that 
one will not obtain the graded Leibniz 
rule in (\ref{ad1})
since the inner product with a multivector field is 
not a graded derivation on the exterior algebra 
but a composition of derivations. 
Nevertheless,  the property following from 
(\ref{ad1}) may be considered as a higher-order analogue of the 
Leibniz rule. 

For example,  for $p=2$ one has $\xx{2}= X^{vi}\der_v\we \der_i$ and
\beqa
&&\xx{2} \inn \dv (\ff{q}\we \ff{r})=
(\xx{2}\inn \dv \ff{q})\we 
\ff{r} + (-1)^q \ff{q} \we (\xx{2} \inn \dv \ff{r}) \nn \\
&& - 2 (-1)^q X^{vi}(\der_v\inn \dv \ff{q}) \we (\der_i \inn \ff{r})
- 2 X^{vi} (\der_i \inn \ff{q}) \we (\der_v\inn \dv \ff{r}) .
\eeqa
This formula may be interpreted as the Leibniz rule for 
the second order graded 
differential operator $\xx{2}$. 
It is similar to the Leibniz rule for 
the second derivative in the usual analysis:
\[ (fg)'' = f'' g + f g'' + 2 f' g' . \]

For an arbitrary $p$ by a straightforward calculation 
we obtain 
\beqa
\xx{p}\inn \dv (\ff{q}\we \ff{r})&=&
\xx{p} \inn\ (\dv \ff{q} \we \ff{r} + (-1)^q 
\ff{q} \we \dv \ff{r}))                      \nn \\
&& \nn \\
&=&(\xx{p} \inn \dv \ff{q}) \we \ff{r} 
+ (-1)^{q(p-1)}\ff{q}
\we (\xx{p} \inn \dv \ff{r}) \nn 
\eeqa
\[ + p(-1)^{p-1}\sum_{s=1}^{p-1} 
\xx{p}{}^v{}\uind{i}{s}{}^{i_{s+1} ... i_{p-1}} 
[(-1)^{q(p-s-1)} 
(\der^{\otimes}_{i_1...i_s}\inn \ff{q})
\we \der^{\otimes}_{i_{s+1} ...  i_{p-1}}\inn\der_v\inn 
\dv \ff{r} \] 
\[ + (-1)^{s(q-p+s-1)} ( \der^{\otimes}_{i_{s+1}...i_{p-1}}
\inn\der_v\inn 
 \dv \ff{q}) \we \der^{\otimes}_{i_1...i_s}\inn \ff{r} ]   \]
where the notation $ \der^{\otimes}_{i_1...i_s} :=
\der_{i_1} \otimes ...
\otimes \der_{i_s} $
is introduced. 
The expression above may be 
viewed as the analogue of the Leibniz rule for the 
graded differential operator of degree $-(p-1)$ and of order $p$.

In terms of the graded Poisson brackets we, therefore, have  
\beqa
&&\pbr{\ff{p}}{\ff{q} \wedge \ff{r}} = 
\pbr{\ff{p}}{\ff{q}} \wedge 
\ff{r} + (-1)^{q(n-p-1)} \ff{q} \wedge \pbr{\ff{p}}{\ff{r}} \nn \\ 
&& \nn \\
&-&(n-p)(-1)^{n-p} 
\xx{n-p}{}^v{}\uind{i}{s}{}^{i_{s+1} ... i_{n-p-1}}
[(-1)^{q(n-p-s-1)}
(\der^{\otimes}_{i_1  ...  i_s}\inn \ff{q}) 
\eeqa
\[ \we \, \der^{\otimes}_{i_{s+1}  ... i_{n-p-1}}\inn\der_v\inn 
\dv \ff{r}  
+ (-1)^{s(n-p-q-s-1)} ( \der^{\otimes}_{i_{s+1}  ...  i_{n-p-1}}
\inn\der_v\inn 
 \dv \ff{q}) \we \der^{\otimes}_{i_1  ... i_s}\inn \ff{r}] \]      
This is the higher-order analogue of the graded Leibniz rule 
for our graded Poisson bracket of forms, eq. (25) in Sect. 3.1.     
A more appropriate formulation in terms of Koszul's $\Phi$ maps 
is given in Sect. 3.1.


\medskip

\medskip

{\bf Acknowledgements.} 
{
\footnotesize 
I gratefully acknowledge the discussions on the earlier 
version of the paper \cite{ikanat-93}  
with 
E. Binz, F. Cantrijn, M. Czachor, M. Gotay, 
J. Kijowski, M. Modugno, Z. Oziewicz, C. Roger, 
G. Rudolph, G. Sardanashvily, 
W. Sarlet, J. S\l awianowski, 
J. \'Sniatycki, I. Tyutin and I. Volovich   
and the letters with useful  comments by P. Michor and M. Crampin.  
I  thank L.A. Dickey for bringing the author's attention to 
his paper \cite{dickey}.
Thanks are due to J. Nester for reading one of the 
earlier versions of the paper and his helpful  comments.  
I also thank J. Hill (Canterbury)  
for his remarks concerning the English 
of an earlier version of the paper. 
It is my pleasure to thank J. S\l awianowski for making 
possible the continuation of this research project at his Laboratory 
and  for his interest, support and benevolence. 
I take the opportunity to express my gratitude to 
Andr\'e Jakob (Aachen) for the help offered to me during a year,  
without which this 
work would not be finished.  
I am also indebted to F. Kim (H\"unxe) for 
his friendly support and aid in 1995. 
My cordial thanks are due to 
M. Pietrzyk (Warsaw) for her  
understanding and invaluable support. 
}





\end{document}